\documentclass[fleqn,10pt]{wlscirep}
\pdfoutput=1
\usepackage{graphicx}
\usepackage{subcaption}  % needed for sub figures
\usepackage{enumerate}
\usepackage{paralist}
\usepackage{xcolor} 

\usepackage{amsmath}
\usepackage{amsfonts}
\usepackage{amssymb}
\usepackage{mathrsfs}  % nice script fonts

\usepackage{bm} % bold math

\newcommand{\red}[1]{{\color{red} #1}}

\newcommand{\cyan}[1]{{\color{cyan} #1}}

\newcommand{\mc}{\mathcal}

\newcommand{\mbf}{\bm} % requires amsmath, amssymb packages

\newcommand{\N}{ \ensuremath{\mathbb{N}}}  % natural numbers
\newcommand{\R}{ \ensuremath{\mathbb{R}}}  % real numbers
\newcommand{\Rn}{\mathbb{R}^{n}}
\newcommand{\cX}{\mathcal{C}}  % configuration space
\newcommand{\cD}{\mathcal{D}}  % domain
\newcommand{\sB}{\mathscr{B}}  % Boolean
\newcommand{\cI}{\mathcal{I}}

\newcommand{\bvec}[1]{\bm{#1}} % uses bm package for boldface math
\newcommand{\x}{\vec{\bvec{x}}}

\newcommand*{\AND}{\land}
\newcommand*{\OR}{\lor}
\newcommand*{\NOT}{\neg}

\newcommand*{\Diff}{\nabla} % differential operator
\newcommand*{\angstrom}{\text{\normalfont\AA}}  % angstrom unit symbol

\newcommand{\norm}[1]{ \ensuremath{\left\lVert{#1}\right\rVert}}
\newcommand{\inlinenorm}[1]{\Vert #1 \Vert}
\newcommand{\ind}[1]{\chi_{#1}}  % indicator function

\newcommand{\multifun}{\mathfrak{m}}

\newcommand{\textbsf}[1]{\textbf{\textsf{#1}}}
\newcommand{\tsf}[1]{\textsf{#1}}

\title{Programmable Potentials: Approximate N-body potentials from coarse-level logic}

\author[1]{Gunjan S. Thakur}
\author[2,*]{Ryan Mohr}
\author[2]{Igor Mezi\'{c}}
\affil[1]{Harvard University, John A.\ Paulson School of Engineering and Applied Sciences, Cambridge, MA 02138, USA}
\affil[2]{University of California Santa Barbara, Department of Mechanical Engineering, Santa Barbara, CA 93106, USA}

\affil[*]{mohrrm@engr.ucsb.edu}

\begin{abstract}
This paper gives a systematic method for constructing an $N$-body potential, approximating the true potential, that accurately captures meso-scale  behavior of the chemical or biological system using pairwise potentials coming from experimental data or ab initio methods. The meso-scale behavior is translated into logic rules for the dynamics. Each pairwise potential has an associated logic function that is constructed using the logic rules, a class of elementary logic functions, and AND, OR, and NOT gates. The effect of each logic function is to turn its associated potential on and off. The $N$-body potential is constructed as linear combination of the pairwise potentials, where the ``coefficients'' of the potentials are smoothed versions of the associated logic functions. These potentials allow a potentially low-dimensional description of complex processes while still accurately capturing the relevant physics at the meso-scale. We present the proposed formalism to construct coarse-grained  potential models for three examples: an inhibitor molecular system, bond breaking in chemical reactions, and DNA transcription from biology. The method can potentially be used in reverse for design of molecular processes by specifying properties of molecules that can carry them out.
\end{abstract}
\begin{document}

\flushbottom
\maketitle
% * <john.hammersley@gmail.com> 2015-02-09T12:07:31.197Z:
%
%  Click the title above to edit the author information and abstract
%
\thispagestyle{empty}

\section*{Introduction}
Multi-body interactions are ubiquitous in nature and happen at all scales from atomic (quantum description) to molecular (classical approach) to macro scales. A systematic analysis these interactions may unfold the fundamental principles governing a given system. For example, understanding the biophysics of protein folding gives insight into disease pathologies \cite{Valastyan:2014gp}. This understanding can be leveraged to develop new vaccines and drug therapies.  Engineering these new products requires accurate and computationally tractable models.

Systems having multibody interactions, in fundamental physics, are often formulated as a ``$N$-body potential'' problem. In order to fully understand these systems a large number of experiments are needed. Conducting experiments may be expensive and at times even impossible. Another approach is to analyze the $N$-body potential governing the system dynamics.  However, at the quantum level, it may be difficult to determine these potentials from first principles due to the complexity of the system. The computational complexity for ab initio methods can scale exponentially in the number of electrons, limiting the practical size of the system to a few thousand atoms \cite{VanDuin:2001ud,Friesner:2005cf,Aktulga:2012dy}. Even if the detailed potential is determined, it may not be immediately useful. Such is the case when the properties or behaviors of interest are at a coarser level than that of the detailed potential and simulating the detailed dynamics is too expensive.
Very coarse approaches such as those of master equation \cite{Gillespie:1992gb} lack predictability on molecular spatial and time scales due to the assumptions with which they are derived. A potential that models the system is required if one is to make predictions about the system.

It is profitable to restrict one's efforts to considering approximate potentials that respect known behavior. Such coarse-level descriptions may be determined from experimental observation and may correspond to trajectories in some  transformed (reaction) coordinate system. For example, consider a signal transduction mechanism \cite{Kiel,Laub,lnui,yarden,sako}, hierarchical self-assembly \cite{berger1,berger2,schwartz,klavins1,klavins2,Whitesides,Whitesides2,yurii,Whitesides3,licata}, Kinesin motor protein translocation on a microtubule \cite{vale,gunjan_prog}, or hydrogen combustion $\mathrm{H_2/O_2}$ \cite{Konnov:2008dy,Hong:2011dj}.  These systems transition from one stable configuration to another on the occurrence of some trigger event which may comprise of an external stimulus or the system reaching a special configuration. 
An external stimulus could be an input of energy that initiates hydrogen combustion, leading to a larger release of energy by the reaction itself.
A special configuration could be a signaling molecule binding to an active receptor site. These stable configurations can be considered as fixed points in a transformed (reaction) coordinate system. The fixed points, the events, and their associated transitions are the coarse-level descriptions that are to be captured in the approximate $N$-body potential.
However, it is still a challenge to construct a $N$-body Hamiltonian potential in a systematic manner that encodes the known coarse-level behaviors into a mathematical formulation and successfully predicts intermediate-scale transition events.

This article introduces a method of encoding coarse-level dynamical behavior into logic functions that are used to ``stitch'' together pairwise interaction potentials into an $N$-body potential. In this method, the practitioner uses experimentally observed coarse-level behavior to derive logic tables that capture various rules of interaction in the system. The qualitative logic tables are turned into a collection of quantitative logic functions associated with pairwise interaction potentials. The logic functions are then turned into smooth encoding functions via a replacement procedure which in turn are used to modify the pairwise potentials. The effect of an encoding function multiplying a pairwise potential is to smoothly turn the potential on or off when a precise set of conditions are met. The combination of the modified potentials gives an $N$-body potential that approximates the true potential governing the system.

\begin{figure}[htbp]
\centering
\includegraphics[width=0.4\columnwidth]{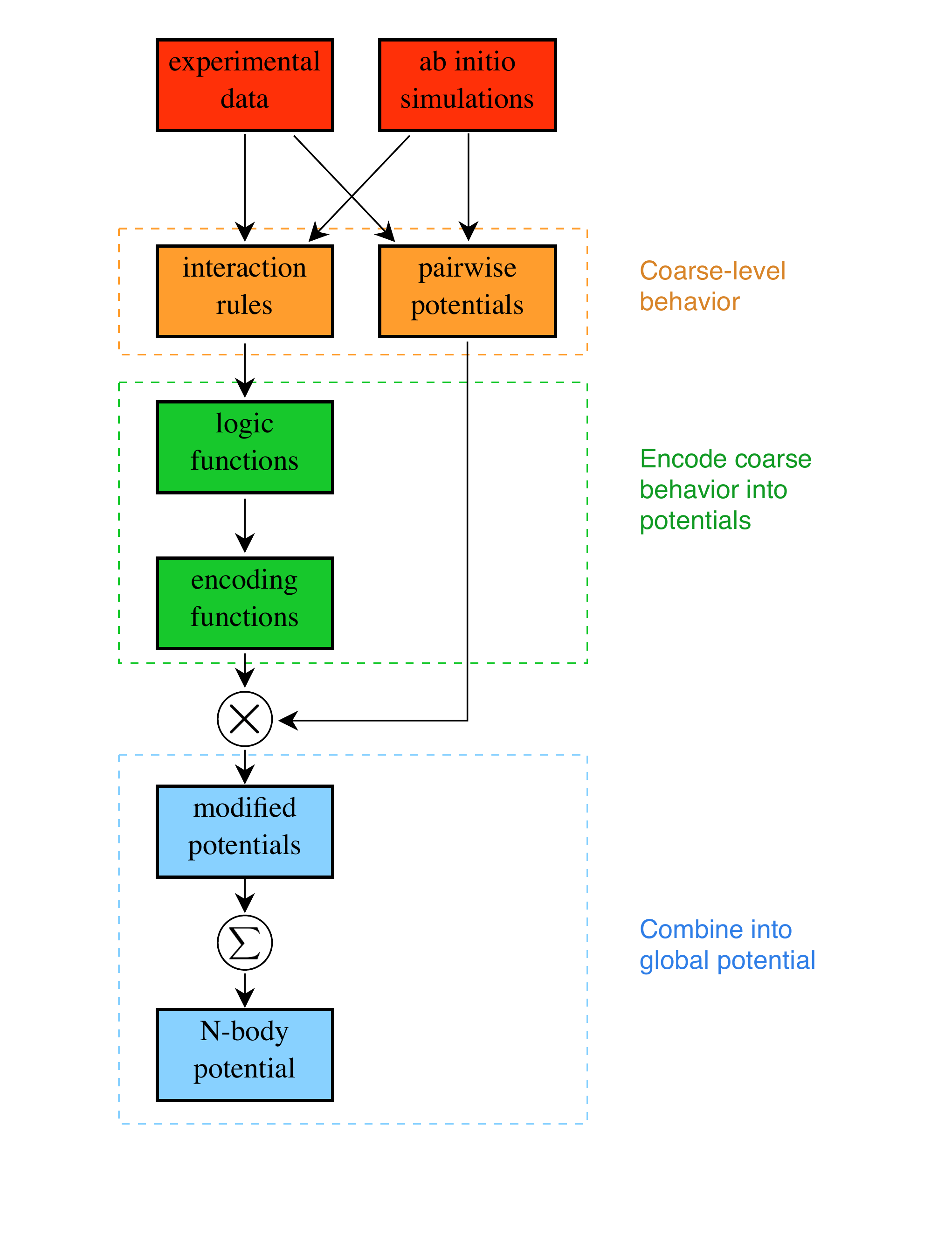}
\caption{Flow chart of procedure. Using experimentally observed data and quantum calculations (red), we extract coarse-grain behavior (orange) namely:  interactions rules and pairwise interaction potentials. This information is used to obtain an N-body potential (blue) for the system by employing the proposed formalism (green).}
\label{fig:flow-chart}
\end{figure}

The method generates a potential that respects what is currently known about the system; it is not claimed that this method results in the unique potential governing the real system. The method does this by leveraging the existing experimental data and the coarse-level behavior that can be derived from it. If more experimental data becomes available, the same procedure can be used to generate a new potential that better models the system. This is equivalent to a refinement of the logic functions and ultimately a refinement of the generated potential. The resulting potential can have a much smaller dimension than the true potential and still accurately capture the relevant physics.

This article begins with a motivating example which is used as an impetus for our modeling framework. In the Methodology section, we define the major components of the framework --- logic functions, permissible logical operations, and the translation to the associated encoding functions --- and specify how they combine with the pairwise potentials to define the approximate potential.  The procedure is depicted in Fig.\ \ref{fig:flow-chart}. 

The procedure is applied to three examples of increasing complexity to showcase the modeling framework. The first is a simple model of an inhibitor molecule mechanism. It shows how one would go from known coarse-level behavior to an approximate global potential that captures that behavior by explicitly constructing the logic tables, the associated logic functions, and the smooth encoding functions. The inhibitor molecule mechanism has more complicated logic than the motivating example and more effectively demonstrates the modeling procedure.

The second example shows how to model a simple, kinetically controlled, bond breaking chemical reaction using this framework. It shows that bond breaking events, and more generally chemical reactions requiring activation energy, can be naturally modeled in the framework. The general procedure for modeling a bond breaking event and how to account for the activation energy is shown. Furthermore, the derived potential is used with LAMMPS \cite{lammps} to numerically simulate the chemical reaction. By changing the relative dissociation energies, the reaction can be biased in a particular direction.

As opposed to our method, many force fields have trouble capturing bond breaking events\cite{Jensen:2007wr}. An exception is the ReaxFF potential \cite{VanDuin:2001ud} that was developed to model reactions of hydrocarbons. The derivation of ReaxFF is based on using interatomic distances to compute the bond order between two atoms and then using the bond order to obtain the bond energy. Corrections to the bond order are dependent on the valency and the deviation of the uncorrected bond order of an atom with its valency. Corrections to the bond energy, in the form of energy penalties (e.g.\ for over-/under-coordination) are added to get the system energy. This is contrasted with our method where bond weakening and breaking is due to the encoding function which is derived from coarse-level observed behavior.

The final example is a simple model of DNA transcription. It is shown that after the binding of RNA polymerase to the promoter region we can sequentially add the complementary base nucleotides to the DNA strand that is to be transcribed. DNA transcription is a complex process involving the interaction of many different molecules \cite{Cramer:2000eb,Hahn:2004hw}. This example shows that we can model such a complex process with a relatively low-dimensional potential that captures the observed mesoscale behavior. To the authors' knowledge there is no other other current potential accomplishing this task.

%%%%%%%%%%%%%%%%%%%%%%%%%%%%%%%%%%%%%%%%%%%%%%%%%%%%%%%%%%%%%%%%%%%%%%%%%%%%%%%%%%%%%%%%%%%%%%%%
%%% MOTIVATING EXAMPLE
\section*{Motivating Example}\label{sec:motivating-example}

\begin{figure}[ht!]
\centering
\includegraphics[width=0.6\columnwidth]{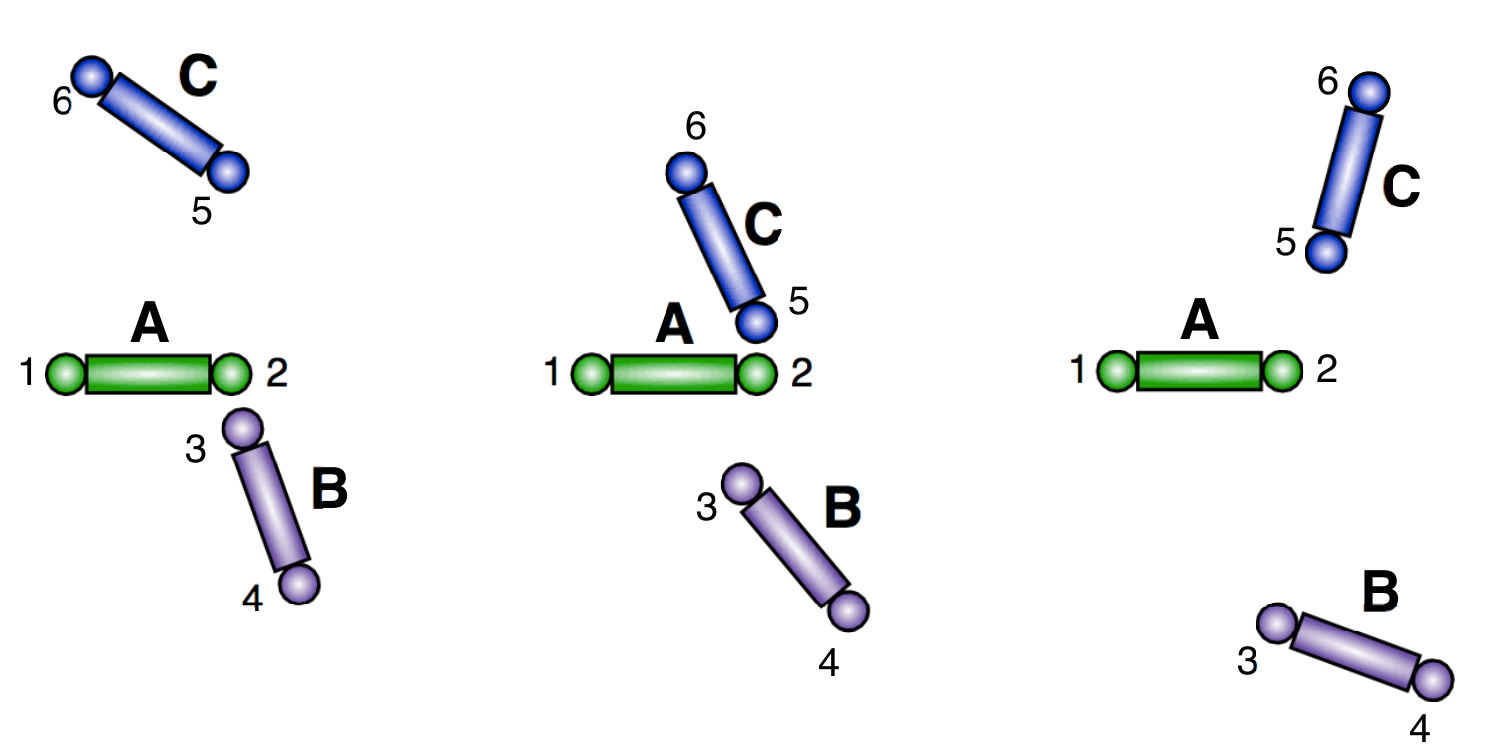}
\caption{Simple signaling molecule mechanism, $\textbsf{A}+ \textbsf{B} \rightarrow \textbsf{AB}\xrightarrow{\textbsf{C}} \textbsf{A} + \textbsf{B}$, modeled by three rods. When \textbsf{C} is not present, \textbsf{A} and \textbsf{B} form a complex. 
When \textbsf{C} is present, \textbsf{A} and \textbsf{B} dissociate and diffuse apart. Molecule \textbsf{C} is free to diffuse away from molecule \textbsf{A}. 
This behavior is captured with the following rules. When atom 5 is far from atom 2, the potential between atoms 2 and 3 is on. When atom 5 is close to atom 2, the potential between atoms 2 and 3 is turned off allowing molecules \textbsf{A} and \textbsf{B} dissociate and diffuse apart. Atom 5 can diffuse away from atom 2.}
\label{fig:interacting-rods-dissociation}
\end{figure}

There are a number of examples in biology where chemical reactions occurring  within a cell are initiated  by some signal or stimulus, followed by an ordered sequence of biochemical reactions. Often the term signal transduction is used to refer to such processes.  One such example is the epidermal growth factor (EGF) signaling \cite{yarden,sako}. Motivated by this example, we construct a  hypothetical system to demonstrate how the proposed formulation can be used to construct a Hamiltonian potential for it. Assume a system of three species, \textbsf{A}, \textbsf{B} and \textbsf{C}, has an evolution dictated by the chemical equation 
	\( \textbsf{A}+ \textbsf{B} \rightarrow \textbsf{AB}\xrightarrow{\textbsf{C}} \textbsf{A} + \textbsf{B}. \)
The sequence in which these reactions happen define logical ``interaction rules" used to design the potential. Specifically, these rules are
\begin{inparaenum}[(1)]
\item when molecules \textbsf{A} and \textbsf{B} are close, and \textbsf{C} is far, then \textbsf{A} and \textbsf{B} bond; and
\item If \textbsf{C} approaches the \textbsf{AB} complex, then \textbsf{A} and \textbsf{B} dissociate.
\end{inparaenum}
This mechanism is visualized in Fig.\ \ref{fig:interacting-rods-dissociation}.

Each of the species in this system can be modeled as a rod having two sites of interaction at the end points, namely atoms $\{1,2\}$ on \textbsf{A}, atoms $\{3,4\}$ on \textbsf{B}, and atoms $\{5,6\}$ on \textbsf{C} (Fig.\ \ref{fig:interacting-rods-dissociation}).
Let us write the force field energy for this system. In general, it is composed of the bonded energies formed from the stretch, bending, and torsional terms, the non-bonded van der Waals and electrostatic terms, and the coupling terms \cite{Jensen:2007wr}. We can split the potential as
	%% ------------ EQUATION ------------ %%
	\begin{equation*}
	U = \sum_{i} \sum_{j> i} \Phi_{(i,j)} + \text{higher order terms}
	\end{equation*}
	%% ---------------- END  ---------------- %%
where $\Phi_{(i,j)}$ denotes a pair-wise interaction potential between two atoms $i$ and $j$.  These 2-atom terms encompass the stretching, torsional, van der Waals, and electrostatic terms and the higher order terms include the bending energies and all the $k$-atom ($k\geq 3$) coupling terms.

For this system, the bonded energy terms are composed of the stretch and torsional energies between the atom-atom pairs $(1,2)$, $(3,4)$ and $(5,6)$, which we can group into a term $U_{b}$. Assume that the only non-negligible non-bonded energy terms are the two van der Waals interactions between atoms 2 and 3 and atoms 2 and 5, and the coupling term between atoms 2, 3, and 5. Denoting these three terms by $\Phi_{(2,3)}$, $\Phi_{(2,5)}$, and $\Phi_{(2,3,5)}$, respectively, we get the force field energy of the system as
	%% ------------ EQUATION ------------ %%
	\begin{equation*}
	U_{\text{signaling}} = \Phi_{(2,3)} + \Phi_{(2,5)} + \Phi_{(2,3,5)} + U_{b} 
	\end{equation*}
	%% ---------------- END  ---------------- %%
The inclusion of the 3-atom potential $\Phi_{(2,3,5)}$ is required in order to capture the transition of the pair $(2,3)$ being in a stable (bounded) configuration when atom 5 is not present to being in an unstable (free) configuration in the presence of the signaling atom 5.

While in general it may be hard to get the correct forms for the coupling term and other higher-order terms in the expansion, and thus the full potential, we know from the above observations that the effect of the potentials $\Phi_{(2,5)}$ and $\Phi_{(2,3,5)}$ is to basically to turn off $\Phi_{(2,3)}$ when 5 is close to 2. Rewriting the potential as
	%% ------------ EQUATION ------------ %%
	\begin{equation*}
	U_{\text{signaling}} =  \Phi_{(2,3)}\left( 1 + \frac{ \Phi_{(2,5)} + \Phi_{(2,3,5)}}{\Phi_{(2,3)}} \right) + U_{b},
	\end{equation*}
	%% ---------------- END  ---------------- %%
this means that term in parentheses is approximately 1 whenever atoms 2 and 5 are far and approximately 0 whenever atoms 2 and 5 are close. Instead of attempting to find the exactly functional forms of $\Phi_{(2,5)}$ and $\Phi_{(2,3,5)}$, we approximate the potential as
	%% ------------ EQUATION ------------ %%
	\begin{equation}\label{eq:potential-approximation-idea}
	U_{\text{signaling}} \approx  S_{(2,3)}\Phi_{(2,3)} + U_{b} ,
	\end{equation}
	%% ---------------- END  ---------------- %%
where $S_{(2,3)}$ is an encoding function that acts as a switch turning $\Phi_{(2,3)}$ on and off. In this example, the encoding function is only function the distance between atoms 2 and 5. The encoding function takes values between 0 and 1, it is approximately 0 when atoms 2 and 5 are close, and approximately 1 when atoms 2 and 5 are far; thus it encodes the logic of the coarse-level observed behavior of the system. It is an approximation of the other terms:
	%% ------------ EQUATION ------------ %%
	\begin{equation*}
	S_{(2,3)} \approx 1 + \frac{ \Phi_{(2,5)} + \Phi_{(2,3,5)}}{\Phi_{(2,3)}}.
	\end{equation*}
	%% ---------------- END  ---------------- %%

In the rest of the article, we make this approximation idea (eq.\ \eqref{eq:potential-approximation-idea}) precise and derive approximate $N$-body potentials from simple pairwise interactions that respect observed coarse-level behavior. We give a systematic procedure to construct the encoding functions which allows us to handle systems with more complex interaction rules. 
We will demonstrate the procedure with three examples. We also use molecular dynamics simulations using the derived potentials that show we can accurately capture the relevant physics.

There are a few items to keep in mind as motivation for the abstract concepts to follow. The basic building blocks for the $N$-body potential are pairwise interaction potentials (denoted by $\Phi_{(2,3)}$ and $\Phi_{(2,5)}$ for the above example). The explicit form of these potentials can be inferred from the experimental data or ab initio reasoning. We approximate the effect of the un-modeled potentials by modifying the relevant pairwise potentials with an encoding function. The encoding function only turns the corresponding potential on and off.  The functional form of the potential  does not change; it is only scaled between 0 and 1. The logic contained in the encoding functions is obtained from experimental observations or ab initio simulations and the logic only depends on pairwise distances between particles, except the pairwise distance in the argument of the encoding function's corresponding potential function; e.g. the logic corresponding to $\Phi_{(2,3)}$ cannot depend on the distance between atoms 2 and 3.

%%%%%%%%%%%%%%%%%%%%%%%%%%%%%%%%%%%%%%%%%%
%%% MATH FRAMEWORK
\section*{Methodology}\label{sec:math-framework}
It is assumed that there are $M$ interacting entities in a domain $\cD$, where $\cD \subseteq \Rn$, for $n=1,2$, or $3$. Each entity is modeled by a finite number of particles with constraint forces between the particles; the totality of these particles over all the entities are labeled from 1 to $N$. This allows us to treat point particles as well as rigid and and non-rigid bodies. The configuration space is $\cX = \cD^N$. An element, $\x \in \cX$, takes the form $\x = (\bvec{x}_{1}, \dots, \bvec{x}_{N} )$, where $\bvec{x}_{j} \in \cD$ describes the position of particle $j$ in the domain $\cD$.

The dynamics of the system is driven by a potential gradient and external forces. Specifically, the functional form of the dynamics is
	%% ------------ EQUATION ------------ %%
	\begin{equation}\label{eq:system-dynamics}
	m_{i} \ddot{\bvec{x}}_{i} = -\Diff_{\bvec{x}_{i}} \left[ \sum_{\bvec{p} \in \cI} \sum_{j \in \multifun(\bvec{p})} S_{\bvec{p},j}(\x) \Phi_{\bvec{p},j}(\x) \right] + F_{i}(\x,t),
	\end{equation}
	%% ---------------- END  ---------------- %%
where
	%% ------------ EQUATION ------------ %%
	\begin{equation}
	U(\x) = \sum_{\bvec{p} \in \cI} \sum_{j \in \multifun(\bvec{p})} S_{\bvec{p},j}(\x) \Phi_{\bvec{p},j}(\x)
	\end{equation}
	%% ---------------- END  ---------------- %%
is the approximating potential.  The notation in this equation needs to be explained. The set $\cI \subseteq \{1,\dots,N\} \times \{1,\dots, N\}$ is the set that defines which pairs of particles interact via some potential function. For example, if $(2,3)$ is in $\cI$, then there exists a pairwise potential between particles 2 and 3. The multiplicity function $\multifun : \cI \to \N$ determines the repetition of distinct elements of $\cI$; the pair $\mc M = (\cI , \multifun)$ is called a multiset and an element $\bvec{p} \in \cI$ appears $\multifun(\bvec{p})$ times in $\mc M$. We make the convention that the pairs $(i,j)$ and $(j,i)$ define the the same element of the multiset. Often $\multifun(\bvec{p})$ will be 1 for every element of $\cI$, however, non-unit values become important when a pair $\bvec{p} \in \cI$ interacts through multiple different potentials, each with its own encoding function. Such a non-unit multiplicity functions is useful when modeling, for example, bond-breaking chemical reactions. Two atoms interact through one potential, the bond potential, when they form a stable pair; when this bond is broken, another potential is required to model the electron-electron repulsion between the atoms.
The term $m_{i}$ denotes the mass of particle $i$ and $\Diff_{\bvec{x}_{i}}$ is the gradient operator in the configuration space $\cX$ with respect to the the position $\bvec{x}_{i}$ of particle $i$. The term $F_{i}(\x)$ collects the external forces on particle $i$. The external force not only contains the forces derived from any external fields, such as electric or magnetic fields, but can also be used to include stochastic effects or boundary constraints. 
 The function $\Phi_{\bvec{p},j} : \cX \to \R$ denotes the $j^{th}$ potential that the pair $\bvec{p} = (p_1, p_2) \in \cI$ interacts through. The index $j$ is an element of the set $\{1,\dots, \multifun(\bvec{p})\}$; i.e., it is the $j^{th}$ instance that pair $\bvec{p}$ appears in the multiset.  For $\x = (\bvec x_1, \dots, \bvec x_N)$, it takes the form
	%% ------------ EQUATION ------------ %%
	\begin{equation}\label{eq:pairwise-interaction-potential}
	\Phi_{\bvec{p},j}( \x ) = \phi_{\bvec{p},j} \left( \inlinenorm{ \pi_{p_{1}}(\x) - \pi_{p_{2}}( \x ) } \right) 
	\equiv \phi_{\bvec{p},j} \left( \inlinenorm{ \bvec x_{p_1} - \bvec x_{p_2}  } \right),
	\end{equation}
	%% ---------------- END  ---------------- %%
where for every $i \in \{1,\dots,N\}$, the coordinate map $\pi_{i} : \cX \to \cD$ is linear and extracts the $i^{th}$ element, $\bvec x_i$, of a configuration vector $\x = (\bvec x_1, \dots, \bvec x_N)$; the norm $\inlinenorm{\cdot}$ denotes the normal Euclidean norm; and $\phi_{\bvec{p},j} : \R^{+} \to \R$ is the specific 1D pairwise interaction potential through which the $j^{th}$ copy of pair $\bvec{p}$ interacts.  This potential could, for example, be a Lennard-Jones potential; it can also be different for different interaction pairs. The form given for the potential shows that it is only a function of the distance between $\bvec x_{p_1}$ and $\bvec x_{p_2}$. The function $S_{\bvec{p},j} : \cX \to \R$ in equation \eqref{eq:system-dynamics} is the encoding function, which will be defined shortly, corresponding to the potential $\Phi_{\bvec{p},j}$; it is the encoding function corresponding to the $j^{th}$ potential that atom pair $\bvec{p}$ interacts through. It encodes the coarse-level interaction rules and it is a function of pairwise distances between particles, except for the particle pair $\bvec{p}$ to which it corresponds.  That is, for the interaction pair $\bvec{p} = (p_1, p_2) \in \cI$, the encoding function $S_{\bvec{p},j}$ is \emph{not} a function of the distance $\inlinenorm{\bvec{x}_{p_{1}} - \bvec{x}_{p_{2}}}$. Since the encoding functions and potentials are functions of relative distances only, equation \eqref{eq:system-dynamics} defines a Hamiltonian system \cite{arnold} when we neglect the forces $F_{i}(\x)$.

A majority of the rest of the paper develops the encoding functions and their properties and shows how one would go from coarse-level interaction rules to encoding functions using a few examples. It is assumed that the coarse-level, interaction rules and the interaction potentials $\phi_{j} : \R^+ \to \R$ are known. These come from analyzing experimental data or ab initio simulations and are thus application specific and beyond the scope of this article. Ultimately, the encoding function $S_{\bvec{p},j}$ will be a smoothed version, which is made precise later, of a logic function $L_{\bvec{p},j} : \cX \to \{0,1\}$. The logic function $L_{\bvec{p},j}$ will be the constructed from a finite number of logical operations applied to elementary logic functions from a Boolean algebra. More precisely, the logic function will be an element of the Boolean sub-algebra generated by elementary logic functions. Thus to define the logic functions, it is required to know the specific definitions of the logical operators AND, OR, and NOT (symbolically denoted, $\AND$, $\OR$, $\NOT$) and what Boolean functions are used as the elementary logic functions.

Before getting into the specific definitions, we would like to discuss what properties the logic functions should have and the motivation behind them. As hinted at above, the logic function $L_{\bvec{p},j}$ encodes the coarse-level interaction rules. Its only effect is to turn the corresponding potential $\Phi_{\bvec{p},j}$ on and off depending on specific conditions met by the current configuration of the system. This means that $L_{\bvec{p},j}$ has the effect of multiplying $\Phi_{\bvec{p},j}$ by either 0 or 1. As in the motivating example above, the logic function is only a function of pairwise distances like $\inlinenorm{\bvec{x}_i - \bvec{x}_k}$. However, we choose to restrict $L_{\bvec{p},j}$ so that it is \emph{not} a function of $\inlinenorm{\bvec{x}_{p_1} - \bvec{x}_{p_2}}$. Finally, it would be undesirable for $L_{\bvec{p},j}$ to oscillate wildly when perturbing $\x$. Since $L_{\bvec{p},j}$ takes only the values 0 and 1, it  will have discontinuities in general. The first property we wish $L_{\bvec{p},j}$ to possess is that the set of points of discontinuity is ``small'' in the sense that its complement is open and dense in $\cD$. This means that for most configurations $\x$, if $\x$ is perturbed slightly to $\vec{\bvec{y}}$, then $L_{\bvec{p},j}(\vec{\bvec{y}}) = L_{\bvec{p},j}(\x)$. 
The second property we wish the logic function to possess is similar but not equivalent to the first. Using the notation, $r_{\bvec{p}} = \inlinenorm{\bvec{x}_{p_1} - \bvec{x}_{p_2}}$, we can consider the logic function $L_{\bvec{p},j}$ to be a function of the pairwise distances $r_{\bvec{q}}$ for $\bvec{q} \neq \bvec{p}$. If we fix all the $r_{\bvec{q}}$'s except one, we wish for the resulting 1D function to make only finitely many transitions between 0 and 1. With these goals in mind, precise definitions of logic functions and elementary logic functions can be given.

%%% DEFINE LOGIC FUNCTIONS
A function $b : \cX \to \{0,1\}$ is called a Boolean function on $\cX$ and the set of such all such functions on $\cX$ is denoted as $\sB$. It is easy to see that the functions that are identically 1 and 0 on $\cX$ are Boolean functions. On $\sB$, define for all $f,g \in \sB$ the two binary logical operations AND ($\land$) and OR ($\lor$) and the unary logical operation NOT ($\lnot$) by 
	%% ------------ EQUATION ------------ %%
	\begin{align*}
	(f \land g)(\x) &= f(\x)g(\x), &
	(f \lor g)(\x) &= f(\x) + g(\x) - (f\land g)(\x), &
	(\lnot f)(\x) &= 1 - f(\x). 
	\end{align*}
	%% ---------------- END  ---------------- %%
It can be shown that $(\sB, \lor,\land,1,0)$ is a Boolean algebra with addition and multiplication given by $\lor$ and $\land$, respectively. These three logical operations will be applied to specific elements of the set of all Boolean functions $\sB$ on $\cX$ to generate a Boolean sub-algebra. The logic functions $L_{\bvec{p},j}$ will be elements of this sub-algebra.

The elementary logic functions generating the Boolean sub-algebra are formed from elements of the class of proximity functions. A proximity function $P : \R^+ \to \{0,1\}$ is a non-increasing function. What form do such functions have and why are they used as the elementary logic functions? Every proximity function has the form $P_{R}(r) = \ind{[0,R)}(r)$, for some $R$ satisfying $0\leq R\leq \infty$. The function $\ind{[0,R)} : \R^+ \to \R$ is the indicator function for the semi-open interval $[0,R)$ which takes the value 1 if the argument satisfies $0 \leq r < R$ and 0 otherwise. Note that the functions that are identically 1 or 0 are proximity functions. The \emph{elementary logic functions} are defined as compositions of a proximity function with the coordinate functions $\pi_{i}$ from above. Specifically, the elementary logic function $\ell_{\bvec{q},R}$ for particle pair $\bvec{q} = (q_{1}, q_{2}) \in \cI$ and parameter $0 \leq R \leq \infty$ has the form 
	%% ------------ EQUATION ------------ %%
	\begin{equation}\label{eq:elementary-logic-function}
	\ell_{\bvec{q},R}(\x) 
	= P_{R}(\inlinenorm{\pi_{q_{1}}(\x) - \pi_{q_{2}}(\x) }) 
	\equiv \ind{[0,R)}(\inlinenorm{\bvec{x}_{q_{1}} - \bvec{x}_{q_{2}} }).
	\end{equation}
	%% ---------------- END  ---------------- %%
This function is 1 when $\bvec{x}_{q_{1}}$ and $\bvec{x}_{q_{2}}$ are closer than distance $R$ and 0 when not.

A logic function $L_{\bvec{p},j}$ is generated by applying finitely many of the logical operations $\land$, $\lor$, and $\lnot$ to the elementary logic functions. Recalling from above the desired properties for the logic functions, each $L_{\bvec{p},j}$ corresponding to $\Phi_{\bvec{p},j}$ is only formed from finitely many logical operations on elementary logic functions $\ell_{\bvec{q},R}$ where $\bvec{q} \neq \bvec{p}$. Thus $L_{\bvec{p},j}$ is a function of possibly all the pairwise distances $r_{\bvec{q}} = \inlinenorm{\bvec{x}_{q_1} - \bvec{x}_{q_2}}$ except $r_{\bvec{p}} = \inlinenorm{\bvec{x}_{p_1} - \bvec{x}_{p_2}}$.

%%% LOGIC FUNCTION PROPERTIES
As we desired, each logic function $L_{\bvec{p},j}$ is continuous almost everywhere in $\cX$. This follows since it is composed from pairwise products and sums of elementary logic function, which themselves are continuous almost everywhere. Since the logic function takes the value 0 or 1, this implies that around each point of continuity, there is a neighborhood on which $L_{\bvec{p},j}$ is constant, as was desired. The second desired property follows from the constructing the logic function using a finite number of elementary logic functions.

\begin{figure}[htbp]
\centering
\begin{subfigure}[htbp]{0.49\columnwidth}
	\centering
	\includegraphics[width=1\columnwidth]{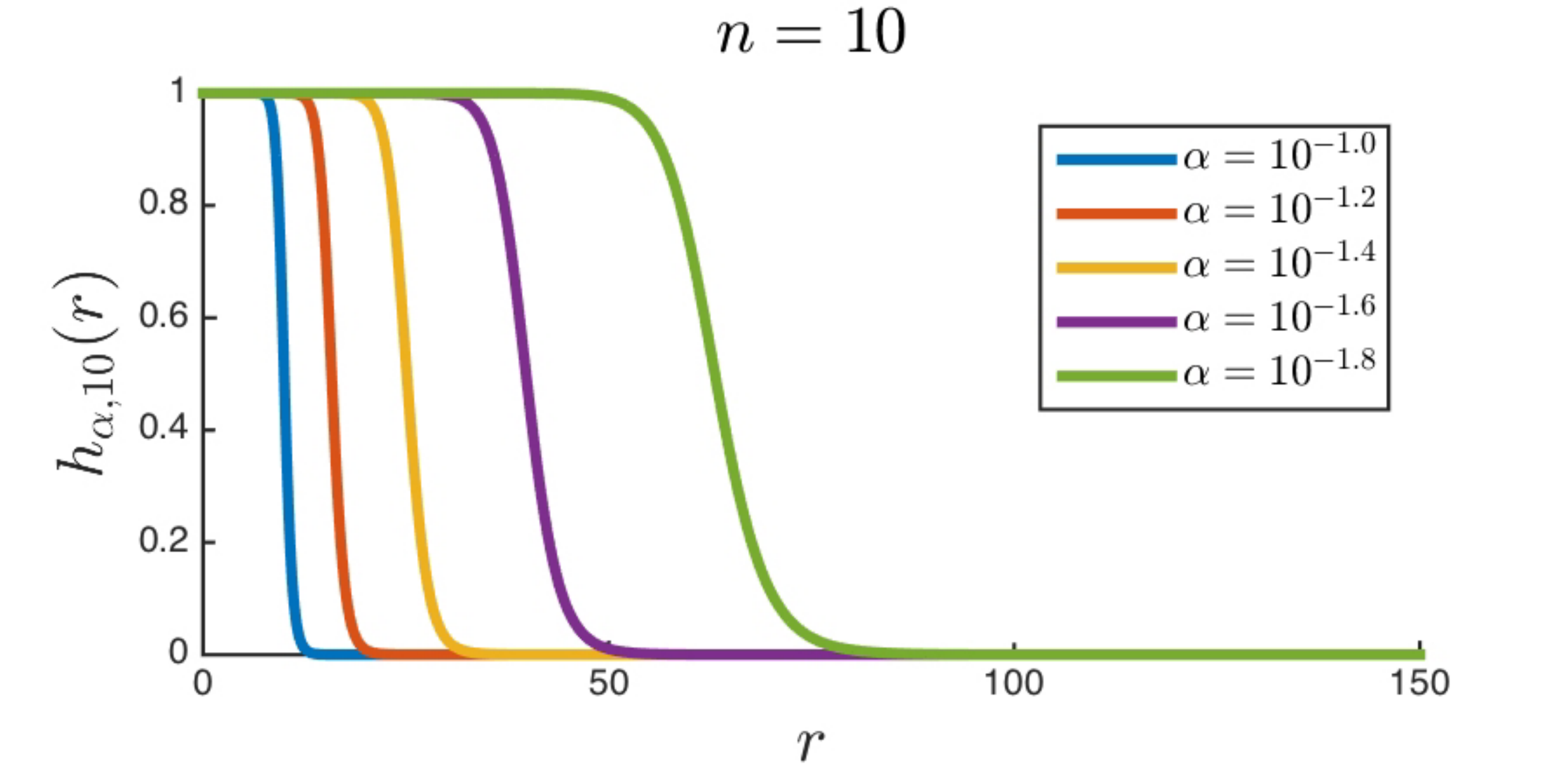}
	\caption{$\alpha$ controls the transition point.}
	\label{fig:h_function_alpha_varied}
\end{subfigure}
\begin{subfigure}[htbp]{0.49\columnwidth}
	\centering
	\includegraphics[width=1\columnwidth]{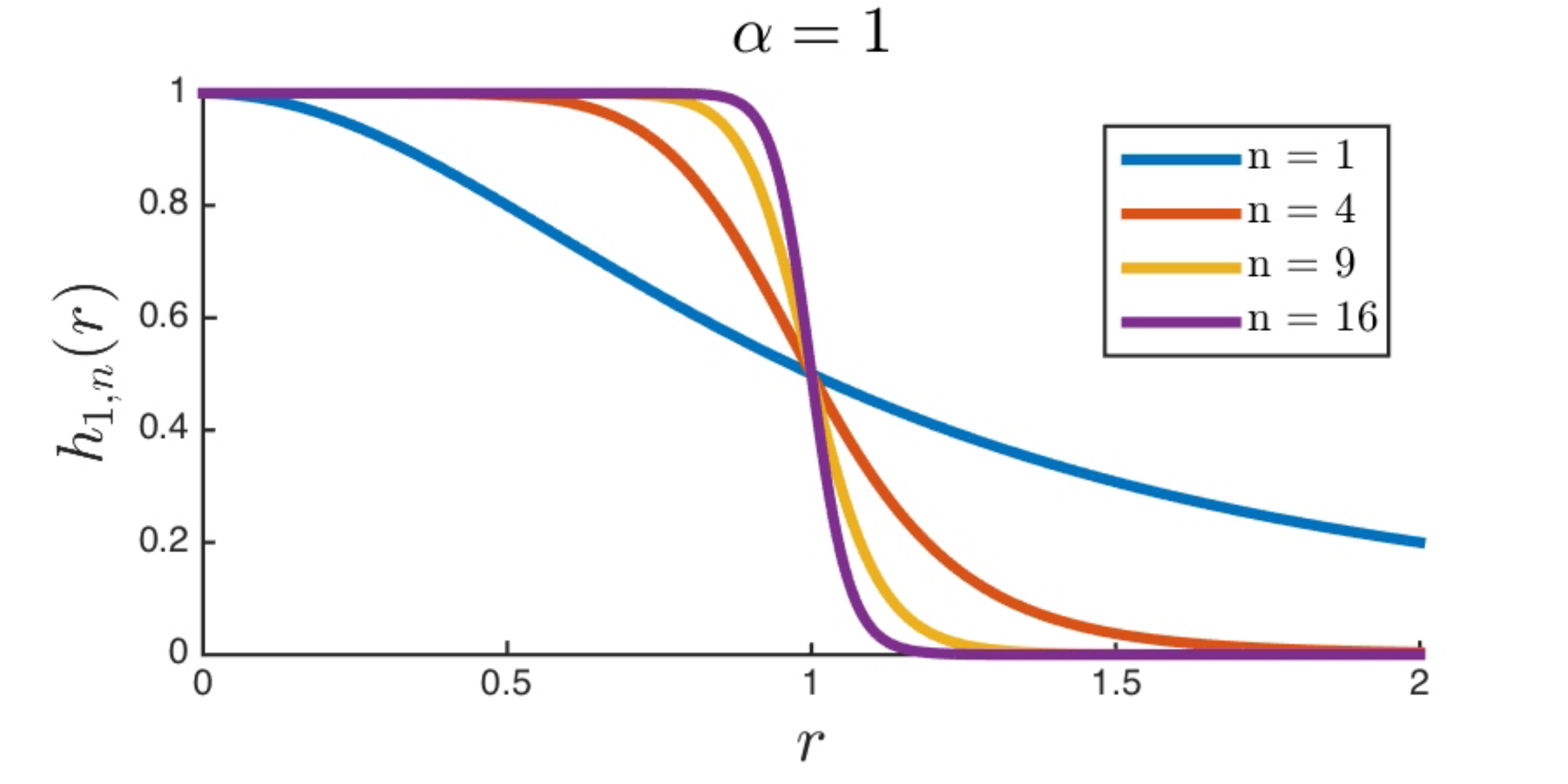}
	\caption{$n$ controls the sharpness of the transition}
	\label{fig:h_function_n_varied}
\end{subfigure}
\caption{Behavior of the $h_{\alpha,n}$ function from equation \eqref{eq:h-function}}
\label{fig:h_function_variation}
\end{figure}

%%% TRANSLATE LOGIC FUNCTIONS TO ENCODING FUNCTIONS
Once the logic function is specified, it must be translated into a smooth encoding function. Ideally, this would be accomplished via a convolution in the ($N$-dimensional) configuration space with a smooth, nonnegative summability kernel (see \cite{Katznelson:2002uy} for a definition). Analytically, this is intractable, and computationally, this is very expensive. Instead, we individually smooth each of the 1D elementary logic functions $\ell_{\bvec{q},R}(\x)$ in the expression for $L_{\bvec{p},j}(\x)$. This is done by replacing the proximity function of $\ell_{\bvec{q},R}$ with a smoothed version. Again, this could be done via the convolution (now 1-dimenional) of each proximity function with a smooth, 1D summability kernel or, alternatively, by the replacement of each indicator function with a specific functional form. In this article, we choose the latter approach and replace each proximity function $\ind{[0,R)}(r)$ with a function of the form
	%% ------------ EQUATION ------------ %%
	\begin{equation}\label{eq:h-function}
	h_{\alpha,n}(r) = \frac{1}{1 + (\alpha r)^{2n}}, \qquad (\alpha \geq 0, n \in \N).
	\end{equation}
	%% ---------------- END  ---------------- %%
For example, if the logic function has the  expression 
	%% ------------ EQUATION ------------ %%
	\begin{equation*}
	L_{\bvec{p},j}(\x) = \ind{[0,R_1)}(\inlinenorm{\bvec{x}_{q_1} - \bvec{x}_{q_2}}) \AND (1 - \ind{[0,R_2)}(\inlinenorm{\bvec{x}_{s_1} - \bvec{x}_{s_2}})),
	\end{equation*}
	%% ---------------- END  ---------------- %%
then the corresponding encoding function would be 
	%% ------------ EQUATION ------------ %%
	\begin{equation*}
	S_{\bvec{p},j}(\x) = h_{\alpha_1,n_1}(\inlinenorm{\bvec{x}_{q_1} - \bvec{x}_{q_2}}) \land (1 - h_{\alpha_2, n_2}(\inlinenorm{\bvec{x}_{s_1} - \bvec{x}_{s_2}})),
	\end{equation*}
	%% ---------------- END  ---------------- %%
for some choices of parameters $\alpha_1, \alpha_2, n_1,$ and $n_2$. The parameters $\alpha$ and $n$ control how well $h_{\alpha,n}$ approximates a proximity function. In particular, $h_{\alpha,n}(0) = 1$ for all nonnegative $\alpha$ and positive $n$. For $\alpha >0$, $\lim_{r\to\infty} h_{\alpha,n}(r) = 0$ and it is strictly monotonically decreasing. Note that for any $n$, $h_{0,n}(r) = 1$ for all $r\geq 0$.  On the other hand, for a fixed $\alpha > 0$, the transition from 1 to 0 becomes sharper as $n$ increases (Fig.\ \ref{fig:h_function_n_varied}). To match a specific indicator function $\ind{[0,R)}$, choose $\alpha = 1/R$. With this choice of $\alpha$, the function satisfies $h_{\frac{1}{R},n}(R) = 1/2$ for any $n$ and as $n \to \infty$,
	%% ------------ EQUATION ------------ %%
	\begin{equation*}
	\lim_{n\to\infty} h_{\frac{1}{R},n}(r) = \begin{cases}
	1 & r < R \\
	\frac{1}{2} & r = R \\
	0 & r > R .
	\end{cases}
	\end{equation*}
	%% ---------------- END  ---------------- %%
%

%

%%%%%%%%%%%%%%%%%%%%%%%%%%%%%%%%%%%%%%
%%% EXAMPLES
%%%%%%%%%%%%%%%%%%%%%%%%%%%%%%%%%%%%%%
\section*{Examples}
To show the entire process, starting from coarse, interaction rules and recovering the encoding function, we apply the method to three examples in increasing order of complexity. The first example is a model for an inhibitor molecule system and is used to exhibit the core methodology of the modeling framework. This system can be considered as an extension of the signaling molecule example above (Fig.\ \ref{fig:interacting-rods-dissociation}). The second example is a model for a simple bond breaking chemical reaction and makes use of the multiplicity function $\multifun$ from the framework. It is shown that the bond dissociation energy is accurately captured in this framework. Numerical simulations show that 
\begin{inparaenum}[(i)]
\item the system exhibits the same coarse-level behavior that was used to derive the potential and 
\item that biased chemical reactions are easily handled. 
\end{inparaenum}
The final example is a simple model for DNA transcription and is the most complicated of the three. This example shows that the logic, and hence potential, of real systems can be captured in the modeling framework in a straight-forward manner.

%%%%%%%%%%%%%%%%%%%%%%%%%%%%%%%%%%%%%%
%%% EXMAPLE 1: INHIBITOR MECHANISM
%%%%%%%%%%%%%%%%%%%%%%%%%%%%%%%%%%%%%%
\subsection*{Simple inhibitor molecule mechanism}\label{subsec:inhibitor-mechanism}
\begin{figure}[htbp]
\centering
\includegraphics[width=0.7\columnwidth]{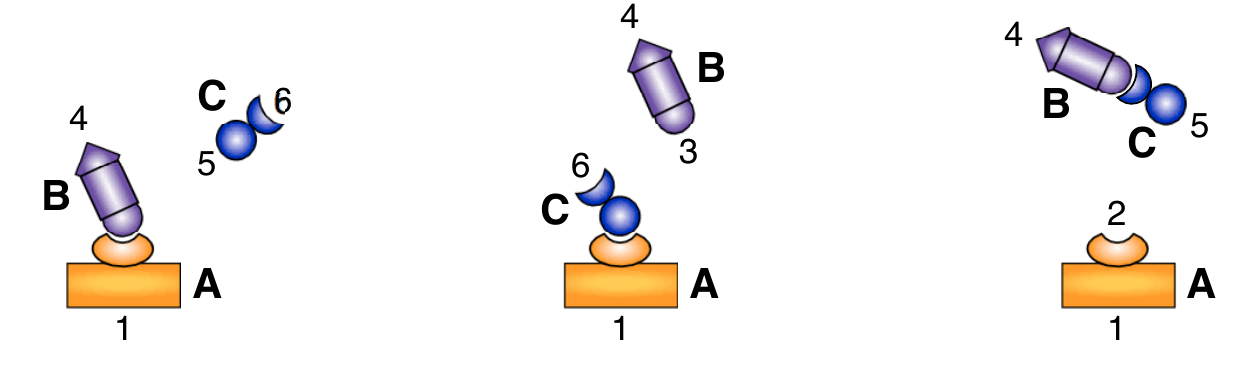}
\caption{Inhibitor molecule example. When the inhibitor molecule, \textbsf{C}, is not present, a bond between receptor \textbsf{A}'s site 2 and site 3 on the active molecule \textbsf{B} can form. When the inhibitor molecule is present, it can either take up the receptor site through a (2,5) bond or bind to site 3 on \textbsf{B} with site 6. Either of these cases prevents to active molecule \textbsf{B} from binding with its receptor site on \textbsf{A}.}
\label{fig:inhibitor-molecule}
\end{figure}

This example can be thought of as a simple model for the action of an inhibitor molecule in a plane. 
Consider the three interacting molecules in Fig.\ \ref{fig:inhibitor-molecule}. The configuration space for this example is $\cX = (\R^2)^6$, where $\x \in \cX$ is written as $\x = (\mbf x_{1}, \dots, \mbf x_{6})$. The set of interacting pairs is $\cI = \{ (2,3), (2,5), (3,6) \}$. For this example, the multiplicity function $\multifun : \cI \to \N$ is identically 1. Thus we have the pairwise potentials $\Phi_{(2,3)}$, $\Phi_{(2,5)}$, and $\Phi_{(3,6)}$. It is assumed that these potentials are formed using a Morse potential (see \eqref{eq:morse-potential}). Molecule \textbsf{C} is an inhibitor molecule and prevents the formation of the $\textbsf{AB}$ complex. Without \textbsf{C}, we have $\textbsf{A} + \textbsf{B} \rightarrow \textbsf{AB}$. With \textbsf{C} present, the there are two possibilities:
\begin{inparaenum}[(i)]
\item $\textbsf{A} + \textbsf{B} \xrightarrow{\textbsf{C}} \textbsf{AC} + \textbsf{B}$ or
\item $\textbsf{A} + \textbsf{B} \xrightarrow{\textbsf{C}} \textbsf{A} + \textbsf{BC}$.
\end{inparaenum}

This behavior is captured in the logic functions $L_{(2,3)}$, $L_{(2,5)}$ and $L_{(3,6)}$. The logic function $L_{(2,3)}$ is 0, i.e., the potential $\Phi_{(2,3)}$ is turned off, when either atom 5 is close to atom 2 or when atom 6 is close to atom 3. This is different from the motivational example which only turned off the potential if 2 and 5 were close and the bonds between 2 and 5 or 3 and 6 never formed. Additionally, if $\textbsf{AC}$ has formed (atoms 2 and 5 close), then $\textbsf{BC}$ cannot form, i.e., $L_{(3,6)} = 0$ and $\Phi_{(3,6)}$ is off. Similarly, $\textbsf{BC}$ has formed (atoms 3 and 6 close), then $\textbsf{AC}$ cannot form. Table \ref{table:inhibitor-logic} captures this logic.
\begin{table}[htbp]%The best place to locate the table environment is directly after its first reference in text
\caption{Bond logic for the inhibitor molecule mechanism.}
\label{table:inhibitor-logic}
\begin{tabular}{cc|c|c|c}
\hline
2 and 5 ``close'' & 3 and 6 ``close'' & $L_{(2,3)}$  & $L_{(3,6)}$  & $L_{(2,5)}$ \\
\hline
0 & 0  & 1  & 1  & 1\\
0 & 1  & 0  & 1  & 0\\
1 & 0  & 0  & 0  & 1\\
1 & 1  & 0  & 0  & 0\\
\hline
\end{tabular}
\end{table}

We need to specify what is meant by ``close''. We assume that ``close'' is in this case is determined by experiments to mean being within the distances $R_{(2,5)}$ and $R_{(3,6)}$, respectively. Thus, atoms 2 and 5 are close when the elementary logic function $\ell_{(2,5), R_{(2,5)}}$ evaluates to 1 and not close when it evaluates to 0.  Using the table, $L_{2,3}(\x)$ corresponding to the interaction potential $\Phi_{(2,3)}(\x)$ can be written as equation \eqref{eq:L_equation_AB}.
	%% ------------ EQUATION ------------ %%
	\begin{align}
	L_{(2,3)}(\x) &= \NOT [\ell_{(2,5), R_{(2,5)}}(\x) \OR \ell_{(3,6), R_{(3,6)}}(\x) ] \nonumber \\
	&=\NOT\left[ \ind{[0,R_{(2,5)})}(\norm{\mbf x_{5}- \mbf x_{2}}) \OR \ind{[0,R_{(3,6)})}(\norm{\mbf x_{6}- \mbf x_{3}}) \right] \nonumber \\
	&= \NOT\left[\ind{[0,R_{(2,5)})}(\norm{\mbf x_{5}- \mbf x_{2}}) + \ind{[0,R_{(3,6)})}(\norm{\mbf x_{6}- \mbf x_{3}}) 
	 - \ind{[0,R_{(2,5)})}(\norm{\mbf x_{5}- \mbf x_{2}})\cdot \ind{[0,R_{(3,6)})}(\norm{\mbf x_{6}- \mbf x_{3}}) \right] \nonumber \\
	&=1 - \ind{[0,R_{(2,5)})}(\norm{\mbf x_{5}- \mbf x_{2}}) - \ind{[0,R_{(3,6)})}(\norm{\mbf x_{6}- \mbf x_{3}}) 
	+ \ind{[0,R_{(2,5)})}(\norm{\mbf x_{5}- \mbf x_{2}})\cdot \ind{[0,R_{(3,6)})}(\norm{\mbf x_{6}- \mbf x_{3}}). \label{eq:L_equation_AB}
	\end{align}
	%% ---------------- END  ---------------- %%
The other logic functions are
	%% ------------ EQUATION ------------ %%
	\begin{align}
	L_{(3,6)}(\x) &= \NOT \ind{[0,R_{(2,5)})}(\norm{\mbf x_{5} - \mbf x_{2}}) \label{eq:L_equation_BC} \\
	L_{(2,5)}(\x) &= \NOT \ind{[0,R_{(3,6)})}(\norm{\mbf x_{6} - \mbf x_{3}}) \label{eq:L_equation_AC}.
	\end{align}
	%% ---------------- END  ---------------- %%

To turn the logic functions into an encoding function, replace each of the proximity functions, $\ind{(0,R_{\bvec{p}}]}$, in \eqref{eq:L_equation_AB} - \eqref{eq:L_equation_AC} with their smooth versions, $h_{1/R_{\bvec{p}},n_{\bvec{p}}}$ (eq.\ \eqref{eq:h-function}). The encoding function $S_{(2,3)}$ corresponding to $L_{(2,3)}$ is
	%% ------------ EQUATION ------------ %%
	\begin{equation}
	S_{(2,3)}(\x)
	= 1 - h_{\alpha_{(2,5)},n_{(2,5)}}(\norm{\mbf x_{5}- \mbf x_{2}})
	- h_{\alpha_{(3,6)},n_{(3,6)}}(\norm{\mbf x_{6}- \mbf x_{3}}) 
	+ h_{\alpha_{(2,5)},n_{(2,5)}}(\norm{\mbf x_{5}- \mbf x_{2}}) h_{\alpha_{(3,6)},n_{(3,6)}}(\norm{\mbf x_{6}- \mbf x_{3}}), \label{eq:EF-function}
	\end{equation}
	%% ---------------- END  ---------------- %%
where $\alpha_{\bvec{p}} = 1/R_{\bvec{p}}$, and similarly for the other two. The approximate $N$-body potential for this system is
	%% ------------ EQUATION ------------ %%
	\begin{equation}\label{eq:potential-inhibitor-molecule}
	U(\x) = \sum_{\bvec{p} \in \cI} S_{\bvec{p}}(\x)\Phi_{\bvec{p}}(\x)  
	= S_{(2,3)}(\x) \Phi_{(2,3)}(\x) + S_{(2,5)}(\x) \Phi_{(2,5)}(\x) 
	+ S_{(3,6)}(\x)\Phi_{(3,6)}(\x) .
	\end{equation}
	%% ---------------- END  ---------------- %%
The original configuration space was 12-dimensional. However, \eqref{eq:potential-inhibitor-molecule} is 8-dimensional since it only depends on four particles (four unique particles making pairs in $\cI$). Thus we were able to reduce the dimension of the configuration space and still capture the relevant physics.

\begin{figure}[htbp]
\begin{center}
\includegraphics[width = 0.5\textwidth]{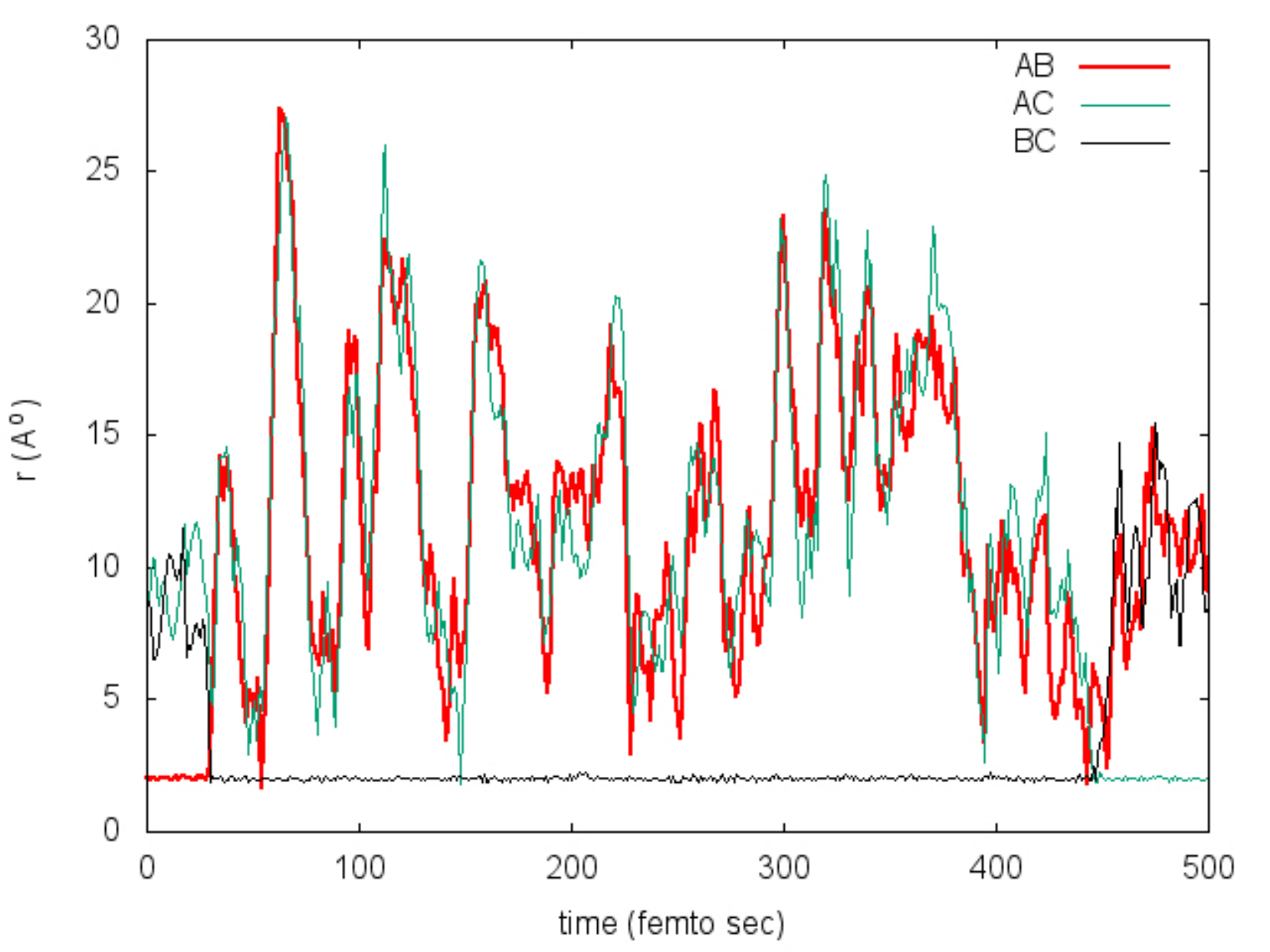}
\caption{One realization of the inhibitor molecule system \eqref{eq:potential-inhibitor-molecule} simulated in LAMMPS. Initially, the \textbsf{AB} complex is formed. Around 30 femtoseconds \textbsf{C} comes close enough, turns off the \textbsf{AB} bond and \textbsf{BC} forms and can diffuse away from \textbsf{A}. This remains the case until around 450 femtosceonds, when \textbsf{A} approaches \textbsf{BC}, the \textbsf{BC} bond turns off and the \textbsf{AC} bond turns on.}
\label{fig:inhibitor-molecule-simulation-trace}
\end{center}
\end{figure}

Here, we are only interested in demonstrating the methodology qualitatively so we make the approximation that derivative of each encoding function is $0$ almost everywhere (this would be the case if the logic functions were used in place of the encoding functions in \eqref{eq:potential-inhibitor-molecule}. With this approximation the force only consists of terms of the form $S_{\bvec p}(\x) \nabla \Phi_{\bvec{p}}(\x)$.
One realization of the inhibitor molecule system \eqref{eq:potential-inhibitor-molecule} simulated in LAMMPS \cite{lammps} with this approximation is shown in Fig.\ \ref{fig:inhibitor-molecule-simulation-trace}. The potentials $\Phi_{\bvec{p}}$ are formed from Morse potentials
	%% ------------ EQUATION ------------ %%
	\begin{equation}\label{eq:morse-potential}
	\phi_{\text{Morse}}(r) = D \left( e^{-2a( r - r^{eq} )} - 2e^{-a( r - r^{eq} )} \right),
	\end{equation}
	%% ---------------- END  ---------------- %%
where $D$ is the dissociation energy, $r^{eq}$ is the equilibrium distance of the bond, and $a$ is a parameter.
Simulations are performed by solving the Langevin equations at constant temperature (i.e.\ NVE ensemble). The parameters used in the computations are given in Supplementary Table I.
Initially, the \textbsf{AB} complex is formed. Around 30 femtoseconds \textbsf{C} comes close enough, turns off the \textbsf{AB} bond and \textbsf{BC} forms and can diffuse away from \textbsf{A}. This remains the case until around 450 fs, when \textbsf{A} approaches \textbsf{BC}, the \textbsf{BC} bond turns off and the \textbsf{AC} bond turns on.

Supplementary Movie 1 shows a simulation of the inhibitor molecule mechanism.

%%%%%%%%%%%%%%%%%%%%%%%%%%%%%%%%%%%%%%%%%%%%%%%%%%%
%%% EXAMPLE 2: BOND BREAKING CHEM REACTION
%%%%%%%%%%%%%%%%%%%%%%%%%%%%%%%%%%%%%%%%%%%%%%%%%%%
\subsection*{Modeling a bond breaking chemical reaction}\label{subsec:chemical-reaction}

Let us consider a slightly more complicated example than the inhibitor molecule. We model a bond breaking, reversible, chemical reaction. In particular, we will model the reaction
	%% ------------ EQUATION ------------ %%
	\begin{equation}\label{eq:3-molecule-chemical-reaction}
	\textbsf{AB} + \textbsf{C} \rightleftarrows \textbsf{AC} + \textbsf{B} .
	\end{equation}
	%% ---------------- END  ---------------- %%
Modeling such chemical reactions is difficult with traditional force field methods since they cannot describe changes in the electronic structure and, thus, are unable to describe bond-breaking, bond-forming, charge transfer, etc., of the system undergoing a reaction \cite{Lin:2007tp,Aktulga:2012dy}. Rather than solving the quantum mechanical equations, we take a coarse-level approach and approximate the bond breaking mechanism with logic functions.

\begin{figure}[htbp]
\centering
\includegraphics[width=0.5\columnwidth]{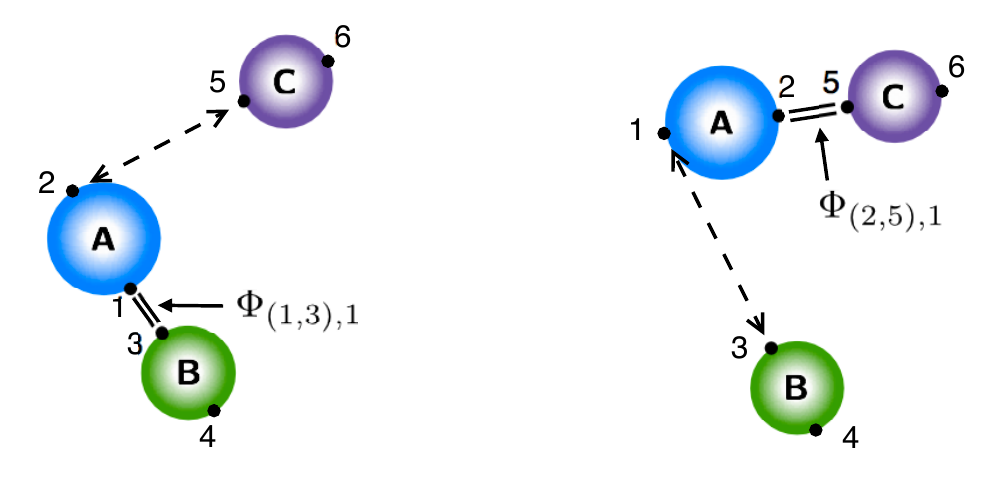}
\caption{Diagram for chemical reaction example. $\Phi_{(1,3),1}$ and $\Phi_{(2,5),1}$ represent the stable bonds \textbsf{AB} and \textbsf{AC}, respectively. The dashed lines represent the repulsion forces induced by the encoding functions. In particular, in the left diagram, the repulsion between \textbsf{A} and \textbsf{C} is due to the partial derivatives of $S_{(1,3),1}(\x)$ with respect to both $\bvec{x}_{2}$ and $\bvec{x}_{5}$. See Supplementary Information Sec.\ III.A for a discussion of the repulsion force induced by the smooth encoding functions.}
\label{fig:chemical-reaction-diagram}
\end{figure}

This example makes use of the multiplicity function $\multifun(\bvec{i})$ in \eqref{eq:system-dynamics} in order to model the electron-electron repulsion during the transition state. It also shows that the use of a smooth encoding function accurately accounts for the bond dissociation energy. Let $\x = (\bvec{x}_{1},\dots, \bvec{x}_{6}) \in (\R^{2})^{6}$ be the configuration vector for this system (see Fig.\ \ref{fig:chemical-reaction-diagram}).

Table \ref{table:chemical-reaction} lists all the potentials involved in modeling the system. The potential $\Phi_{(1,3),1}$ is the potential energy of the bond between \textbsf{A} and \textbsf{B} when they form the stable molecule \textbsf{AB}, for example a Lennard-Jones or Morse potential. Similarly,  $\Phi_{(2,5),1}$ is the bond potential energy between \textbsf{A} and \textbsf{C} when they form the stable molecule \textbsf{AC}. They have the associated logic functions $L_{(1,3),1}$ and $L_{(2,5),1}$, respectively. The potentials $\Phi_{(1,3),2}$ and $\Phi_{(2,5),2}$ are used to model electron-electron repulsion during the transition state when the reactant bonds have broken and the product bonds have not yet formed.

\begin{table}[htbp]%The best place to locate the table environment is directly after its first reference in text
\caption{Potentials in chemical reaction model.}
\label{table:chemical-reaction}
\begin{tabular}{c|c|c|c}
\hline
Potentials & Potential & Equil. & Interaction\\
 & type & dist. & range\\
\hline
$\Phi_{(1,3),1}$ & \textbsf{AB} stable mol. bond & $r_{\textbsf{AB}}^{\textrm{eq}}$ & -- \\
$\Phi_{(1,3),2}$ & \textbsf{AB} electron repulsion term & --  & $<R_{\textbsf{AC},\textbsf{B}}^{\textrm{dis}}$  \\
$\Phi_{(2,5),1}$ & \textbsf{AC} stable mol. bond & $r_{\textbsf{AC}}^{\textrm{eq}}$ & -- \\
$\Phi_{(2,5),2}$ & \textbsf{AC} electron repulsion term & --  & $<R_{\textbsf{AB},\textbsf{C}}^{\textrm{dis}}$  \\
\hline
\end{tabular}
\end{table}
\begin{table}[htbp]%The best place to locate the table environment is directly after its first reference in text
\caption{Bond-breaking logic rules.}
\label{table:chemical-reaction-logic-table}
\begin{tabular}{c|c|c|c|c|c}
\hline
$\norm{\bvec{x}_{1} - \bvec{x}_{3}}$ & $\norm{\bvec{x}_{2} - \bvec{x}_{5}}$ & $\Phi_{(1,3),1}$ & $\Phi_{(1,3),2}$ & $\Phi_{(2,5),1}$ & $\Phi_{(2,5),2}$ \\
\hline
$<R_{\textbsf{AC},\textbsf{B}}^{\textrm{dis}}$ & -- & -- & -- & OFF & ON \\
$>R_{\textbsf{AC},\textbsf{B}}^{\textrm{dis}}$ & -- & -- & -- & ON & OFF \\
-- & $<R_{\textbsf{AB},\textbsf{C}}^{\textrm{dis}}$ & OFF & ON & -- & -- \\
-- & $>R_{\textbsf{AB},\textbsf{C}}^{\textrm{dis}}$ & ON & OFF & -- & -- \\
\hline
\end{tabular}
\end{table}

Table \ref{table:chemical-reaction-logic-table} lists the logic rules for this system. Consider the forward reaction $\textbsf{AB} + \textbsf{C} \rightarrow \textbsf{AC} + \textbsf{B}$. We model the bond breaking mechanism by turning off the stable bond, $\Phi_{(1,3),1}$, when \textbsf{C} gets ``close enough'' to \textbsf{A}. Let $R_{\textbsf{AB},\textbsf{C}}^{\textrm{dis}}$ be the distance of \textbsf{C} to  \textbsf{A} ($\norm{\bvec{x}_{2} - \bvec{x}_{5}}$) within which the bond turns off. Similarly, for the backwards reaction, turn the \textbsf{AC} bond, $\Phi_{(2,5),1}$, off when \textbsf{B} gets within a distance $R_{\textbsf{AC},\textbsf{B}}^{\textrm{dis}}$ of \textbsf{A}. The logic functions are logical NOT's of $\bvec{x}_{2}$-$\bvec{x}_{5}$ and $\bvec{x}_{1}$-$\bvec{x}_{3}$ proximity functions:
	%% ------------ EQUATION ------------ %%
	\begin{align*}
	L_{(1,3),1}(\x) 
	&= \NOT \ind{[0,R_{\textbsf{AB},\textbsf{C}}^{\textrm{dis}})}(\inlinenorm{\bvec{x}_{2} - \bvec{x}_{5}}) \\
	L_{(2,5),1}(\x) 
	&= \NOT \ind{[0,R_{\textbsf{AC},\textbsf{B}}^{\textrm{dis}})}(\inlinenorm{\bvec{x}_{1} - \bvec{x}_{3}}).
	\end{align*}
	%% ---------------- END  ---------------- %%
We assume that $r_{\textbsf{AB}}^{eq} < R_{\textbsf{AC},\textbsf{B}}^{\textrm{dis}}$ and $r_{\textbsf{AC}}^{eq} < R_{\textbsf{AB},\textbsf{C}}^{\textrm{dis}}$.

The use of the smooth encoding function in the potential (as opposed to the logic function) allows the transfer of the correct amount of energy from \textbsf{C} to \textbsf{AB} in order to break the bond; \textbsf{C} must transfer an amount of energy equivalent to the bond dissociation energy $D_{\textbsf{AB}}$ of the \textbsf{AB} bond in order to turn off the $\Phi_{(1,3),1}$ potential. We refer the reader to Sec.\ III.A  in the Supplementary Information for the derivation.
%%% ENERGY TRANSFER CALC IN SUPP INFO

Let us now turn to the case occurring directly after a successful collision of \textbsf{C} with \textbsf{AB}. In this case, \textbsf{A} and \textbsf{B} are close to their equilibrium  distance ($\norm{\bvec{x}_{1} - \bvec{x}_{3}} \approx r_{\textbsf{AB}}^{eq}$) and \textbsf{A} and \textbsf{C} are closer than the \textbsf{AB}-bond dissociation distance ($\norm{\bvec{x}_{2} - \bvec{x}_{5}}  < R_{\textbsf{AB},\textbsf{C}}^{\textrm{dis}}$). In this state, the bonds are weak and neither \textbsf{AB} nor \textbsf{AC} is stable; the system is at its transition state. In this transition state the forces experienced by the molecules due to the bond potentials $\Phi_{(2,3),1}$ and $\Phi_{(2,5),1}$ are small since the encoding functions and their partial derivatives are small, and thus the bond potentials are approximately ``off''. The dynamics are predominantly dominated by noise and the residual momentum of the molecules. 

In this transition state, the electron-electron repulsion should be directly accounted for via a short-range repulsion potential between the molecules, such as by using the repulsive part of a Morse potential. The logic functions are defined such that these repulsion forces are only ``on'' when the system is in its transition state. This is easily accomplished. Denote the short-range repulsion potential between \textbsf{A} and \textbsf{C} by $\Phi_{(2,5),2}$. This force is defined such that $\Phi_{(2,5),2}(\x) \approx 0$, for $\norm{\bvec{x}_{2} - \bvec{x}_{5} } > R_{\textbsf{AB},\textbsf{C}}^{\textrm{dis}}$.  This force is turned on when $\norm{\bvec{x}_{1} - \bvec{x}_{3} } < R_{\textbsf{AC},\textbsf{B}}^{\textrm{dis}}$. The logic function for the \textbsf{A}-\textbsf{C} repulsion is
	%% ------------ EQUATION ------------ %%
	\begin{align}
	L_{(1,3),2}(\x) = 
	\ind{[0,R_{\textbsf{AC},\textbsf{B}}^{\textrm{dis}})}(\norm{\bvec{x}_{1} - \bvec{x}_{3} }). \label{eq:short-range-TS-repulsion}
	\end{align}
	%% ---------------- END  ---------------- %%
Similarly, repulsions between \textbsf{A}-\textbsf{B} and \textbsf{B}-\textbsf{C} can be defined with logic functions similar to the above one.

\begin{figure}[htbp]
\centering
\includegraphics[width=0.5\columnwidth]{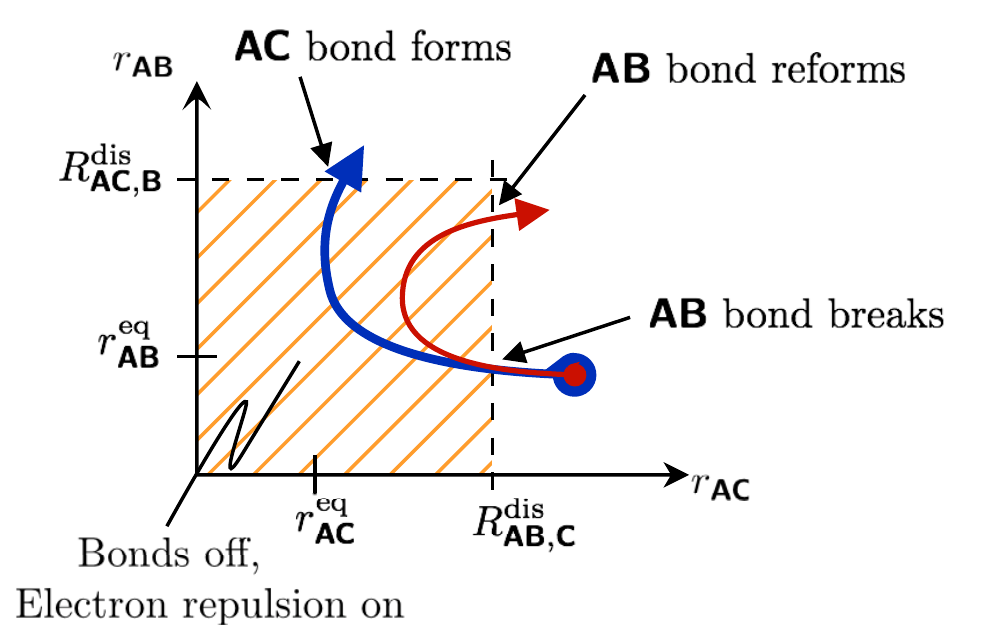}
\caption{The two most probable outcomes of a successful \textbsf{AB} + \textbsf{C} event. Both trajectories start at the same configuration the difference is that \textbsf{C} has a greater momentum for the blue (thicker) curved arrow. For both the red and blue trajectories, \textbsf{C} approaches \textbsf{A}. The \textbsf{AB} bond breaks when $r_{\textbsf{AC}} = \norm{\bvec{x}_{2} - \bvec{x}_{5}} < R_{\textbsf{AB},\textbsf{C}}^{\textrm{dis}}$. Depending on the momentum and the relative strength of the repulsion terms, either \textbsf{AB} reforms or \textbsf{AC} forms. }
\label{fig:chem-reaction-possible-bond-breaking-event}
\end{figure}

There are three possible outcomes for when the system exits its transition state: 
\begin{inparaenum}[(i)]
\item either \textbsf{AC} forms a stable molecule, 
\item \textbsf{AB} reforms, or 
\item no bonds are formed and all the molecules are free molecules.
\end{inparaenum}
 This depends on the equilibrium distances of the bonds, the dissociation distances, the incoming momentum of \textbsf{C}, and the repulsion forces. Figure \ref{fig:chem-reaction-possible-bond-breaking-event} shows the two most probable outcomes for a single $\textbsf{AB} + \textbsf{C}$ event.

%%% BOND BREAKING: NUMERICAL SIMULATIONS
For simulations, the \textbsf{AB} and \textbsf{AC} bonds ($\Phi_{(1,3),1}$ and $\Phi_{(2,5),1}$, respectively) are given by Morse potentials, \eqref{eq:morse-potential}.   In simulations, only the short-range \textbsf{A}-\textbsf{B} and \textbsf{A}-\textbsf{C} electron-electron repulsions are modeled and are only active during the transition state.  The form of these for these repulsions are chosen as the repulsive part of a Morse potential with the same parameters as the full potentials used for the \textbsf{AB} and \textbsf{AC} bonds.  The logic function for the \textbsf{A}-\textbsf{C} repulsion potential, $\Phi_{(2,5),2}$, is given by \eqref{eq:short-range-TS-repulsion} with obvious modifications for $\Phi_{(2,5),2}$. The associated encoding functions are given by the normal replacement procedure. The full potential used during the numerical experiments is given in \eqref{eq:numerical-encoding-potential}.

	\begin{equation}\label{eq:numerical-encoding-potential}
    \begin{aligned}
	U(\x) =& 
	\underbrace{ \left(1-\frac{1}{1+\left(\norm{ \bvec{x}_{2} - \bvec{x}_{5} } / R_{\textbsf{AB},\textbsf{C}}^{\textrm{dis}} \right)^{2n_{\textbsf{AC}}}}\right)}_{S_{(1,3),1}} \underbrace{D_{\textbsf{AB}} \left( e^{-2a(\norm{ \bvec{x}_{1} - \bvec{x}_{3}  } - r_{\textbsf{AB}}^{eq})}  - 2e^{-a(\norm{ \bvec{x}_{1} - \bvec{x}_{3}  } - r_{\textbsf{AB}}^{eq})} \right)}_{\Phi_{(1,3),1}} \\
	&+ \underbrace{ \left(1-\frac{1}{1+ \left(\norm{ \bvec{x}_{1} -\bvec{x}_{3} } /  R_{\textbsf{AC},\textbsf{B}}^{\textrm{dis}}  \right)^{2n_{\textbsf{AB}}}}\right)}_{S_{(2,5),1}} \underbrace{D_{\textbsf{AC}} \left( e^{-2a(\norm{ \bvec{x}_{2} - \bvec{x}_{5}  } - r_{\textbsf{AB}}^{eq})}  - 2e^{-a(\norm{ \bvec{x}_{2} - \bvec{x}_{5}  } - r_{\textbsf{AC}}^{eq})} \right)}_{\Phi_{(2,5),1}} \\
	&+ \underbrace{ \left( \frac{1}{1+\left(\norm{ \bvec{x}_{2}  - \bvec{x}_{5}  } / R_{\textbsf{AB},\textbsf{C}}^{\textrm{dis}} \right)^{2n_{\textbsf{AC}}}} \right) }_{S_{(1,3),2}} 
	\underbrace{ D_{\textbsf{AB}} e^{-2a(\norm{ \bvec{x}_{1} - \bvec{x}_{3}  } - r_{\textbsf{AB}}^{eq})} }_{\Phi_{(1,3),2}} \\
	&+ \underbrace{ \left( \frac{1}{1+ \left(\norm{  \bvec{x}_{1} - \bvec{x}_{3} } /  R_{\textbsf{AC},\textbsf{B}}^{\textrm{dis}}  \right)^{2n_{\textbsf{AB}}}} \right) }_{S_{(2,5),2}} 
	\underbrace{ D_{\textbsf{AC}} e^{-2a(\norm{ \bvec{x}_{2} - \bvec{x}_{5}  } - r_{\textbsf{AB}}^{eq})} }_{\Phi_{(2,5),2}} . 
    \end{aligned}
	\end{equation}

\begin{figure}[htbp]
\centering
\begin{subfigure}[t]{0.3\textwidth}
	\centering
	\includegraphics[width=\textwidth]{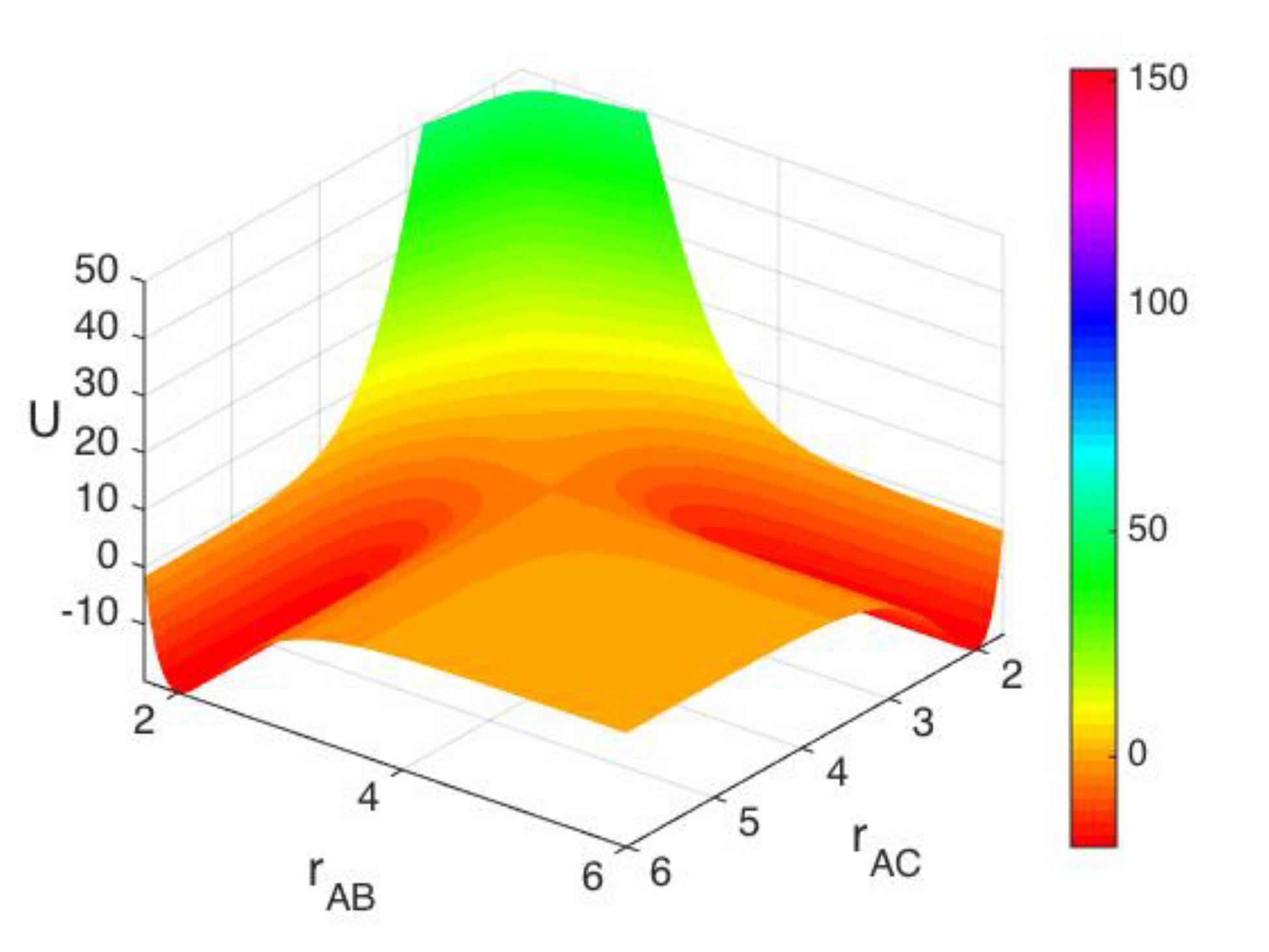}
	\caption{Potential energy surface}
	\label{fig:unbiased-potential-energy-surface}
\end{subfigure}
\begin{subfigure}[t]{0.3\textwidth}
    \centering
    \includegraphics[width=1\textwidth]{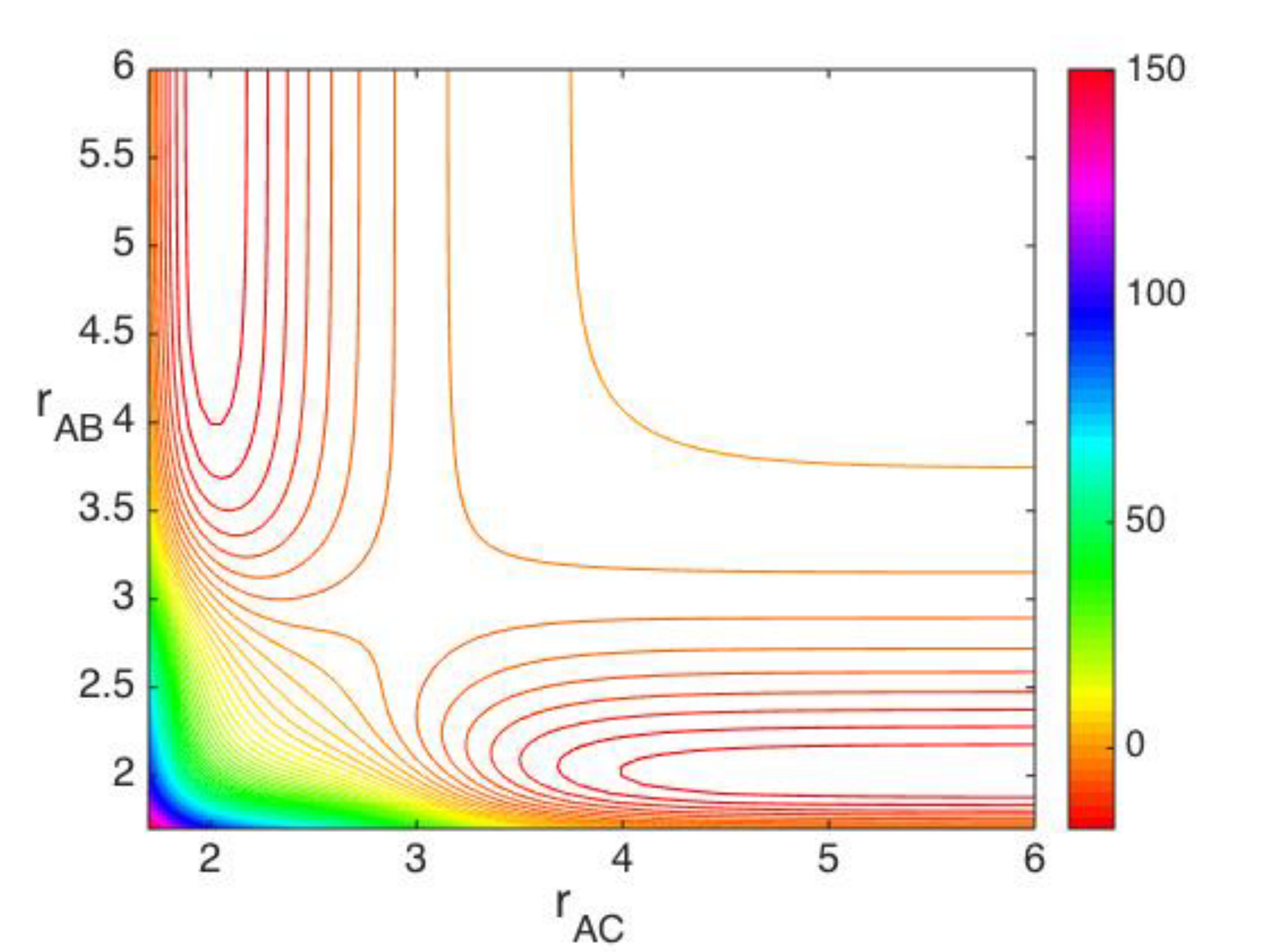}
    \caption{Potential energy level sets}
    \label{fig:unbiased-contour-plot}
\end{subfigure}
\begin{subfigure}[t]{0.3\textwidth}
	\centering
	\includegraphics[width=1\textwidth]{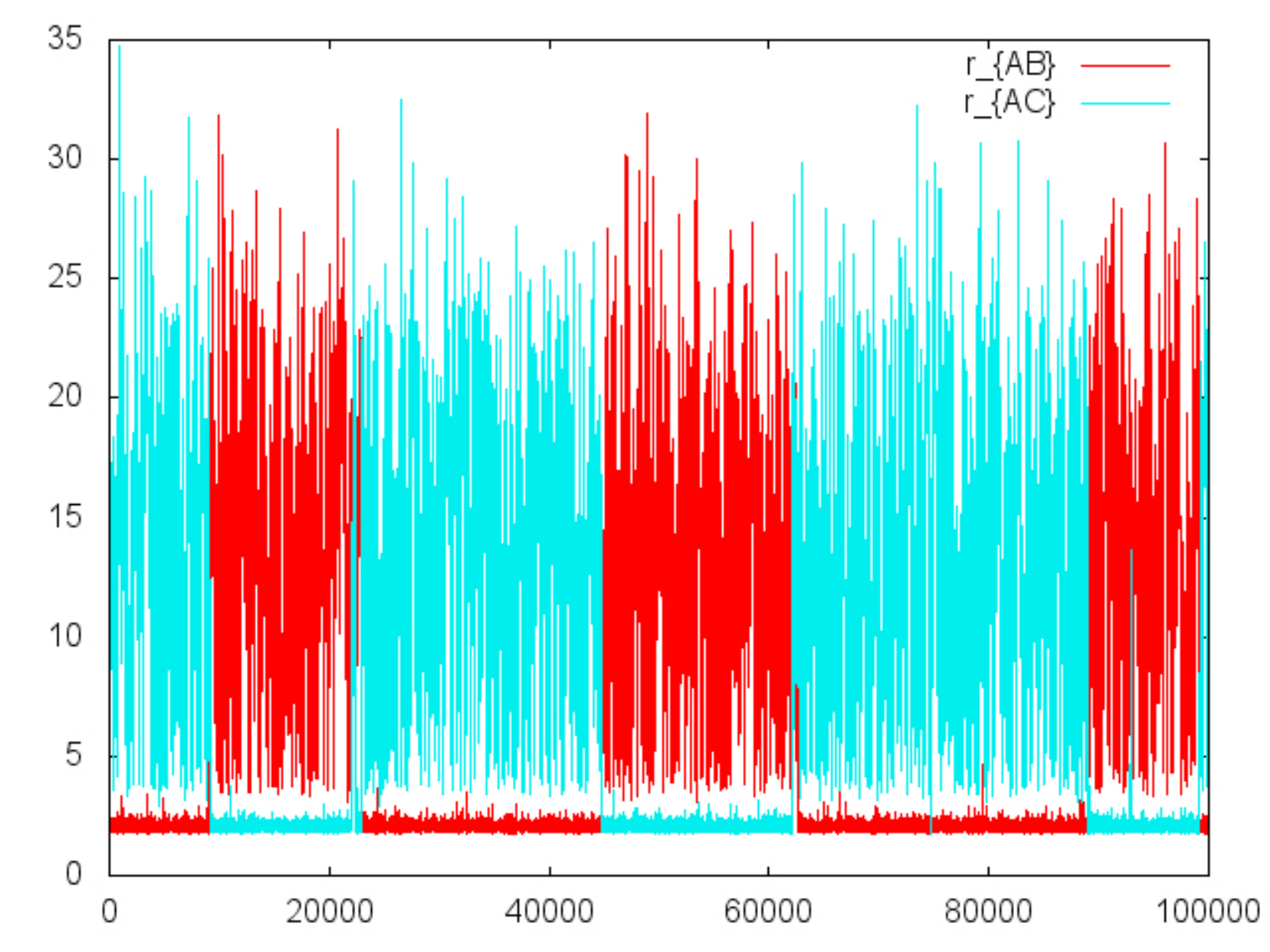}
	\caption{Relative distances between active sites on molecules \textbsf{A}, \textbsf{B} (\red{red}) and molecules \textbsf{A}, \textbsf{C} (\cyan{cyan})}
	\label{fig:unbiased-relative-distances}
\end{subfigure}
\caption{Simulation of an unbiased (1:1 well-depth), bond breaking chemical reaction, \eqref{eq:3-molecule-chemical-reaction}. 
(a) The potential energy \eqref{eq:numerical-encoding-potential} for the system. The parameters are given under simulation 1 in Supplementary Table II. 
(b) The level set plot of the potential energy. 
(c) A typical trajectory of the simulation. The \cyan{cyan} trace denotes the distance between molecules \textbsf{A} and \textbsf{C} ($r_{\textbsf{AC}} = \norm{\bvec{x}_{2} - \bvec{x}_{5}} $), whereas the \red{red} trace corresponds to the distance between molecules \textbsf{A} and \textbsf{B} ($r_{\textbsf{AB}} = \norm{\bvec{x}_{1} - \bvec{x}_{3}} $). Initially, \textbsf{A} and \textbsf{C} are near their equilibrium length (2 \angstrom) and \textbsf{B} is far from \textbsf{A}. We see a successful $\textbsf{AC} + \textbsf{B} \rightarrow \textbsf{AB}  + \textbsf{C}$ event happening very soon (\red{red} trace is close to the equilibrium distance, then becomes large; \cyan{cyan} trace is large, then becomes small).
}
\label{fig:chemical-reaction-simulation-1}
\end{figure}

The force derived from \eqref{eq:numerical-encoding-potential} is used in LAMMPS \cite{lammps} to simulate the system for an unbiased and a biased potential (parameters in Supplementary Information Table II). The parameters of the first simulation are chosen so that the \textbsf{AB} and \textbsf{AC} are symmetric ($D_{\textbsf{AC}}/ D_{\textbsf{AB}} = 1$). In this case, the chemical reaction is unbiased and if averaged over all realizations of the noise, it is expected that the amount of time \textbsf{AB} is formed is equal to the amount of time \textbsf{AC} is formed. Figure \ref{fig:chemical-reaction-simulation-1} shows the potential energy for this simulation (Fig.\ \ref{fig:unbiased-potential-energy-surface}), the corresponding level sets (Fig.\ \ref{fig:unbiased-contour-plot}), and a typical realization of the simulation (Fig.\ \ref{fig:unbiased-relative-distances}). In the energy surface plot and the level set plot, the symmetry of the potential is evident. The realization shown in Fig.\ \ref{fig:unbiased-relative-distances} starts with \textbsf{AB} near its equilibrium length ($2\angstrom$) with \textbsf{C} far from \textbsf{A}. The realization shows the approximately equal times that \textbsf{AB} and \textbsf{AC} are formed. The deviation is due to this being a particular realization rather than an average over an ensemble of realizations and the finite nature of the simulation.

\begin{figure}[htbp]
\centering
\begin{subfigure}[t]{0.3\textwidth}
	\centering
	\includegraphics[width=\textwidth]{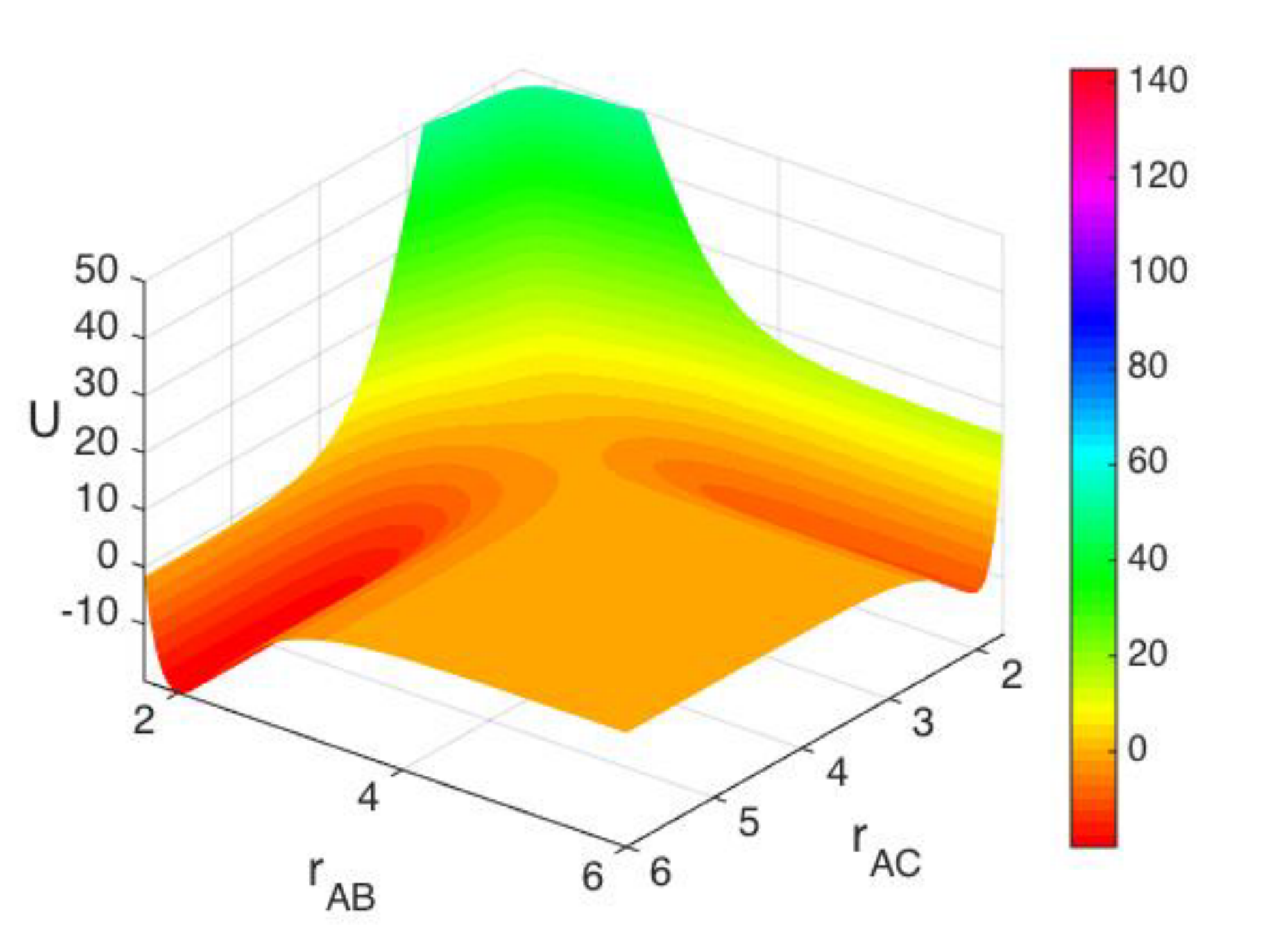}
	\caption{Potential energy surface}
	\label{fig:biased-potential-energy-surface}
\end{subfigure}
\begin{subfigure}[t]{0.3\textwidth}
	\centering
	\includegraphics[width=1\textwidth]{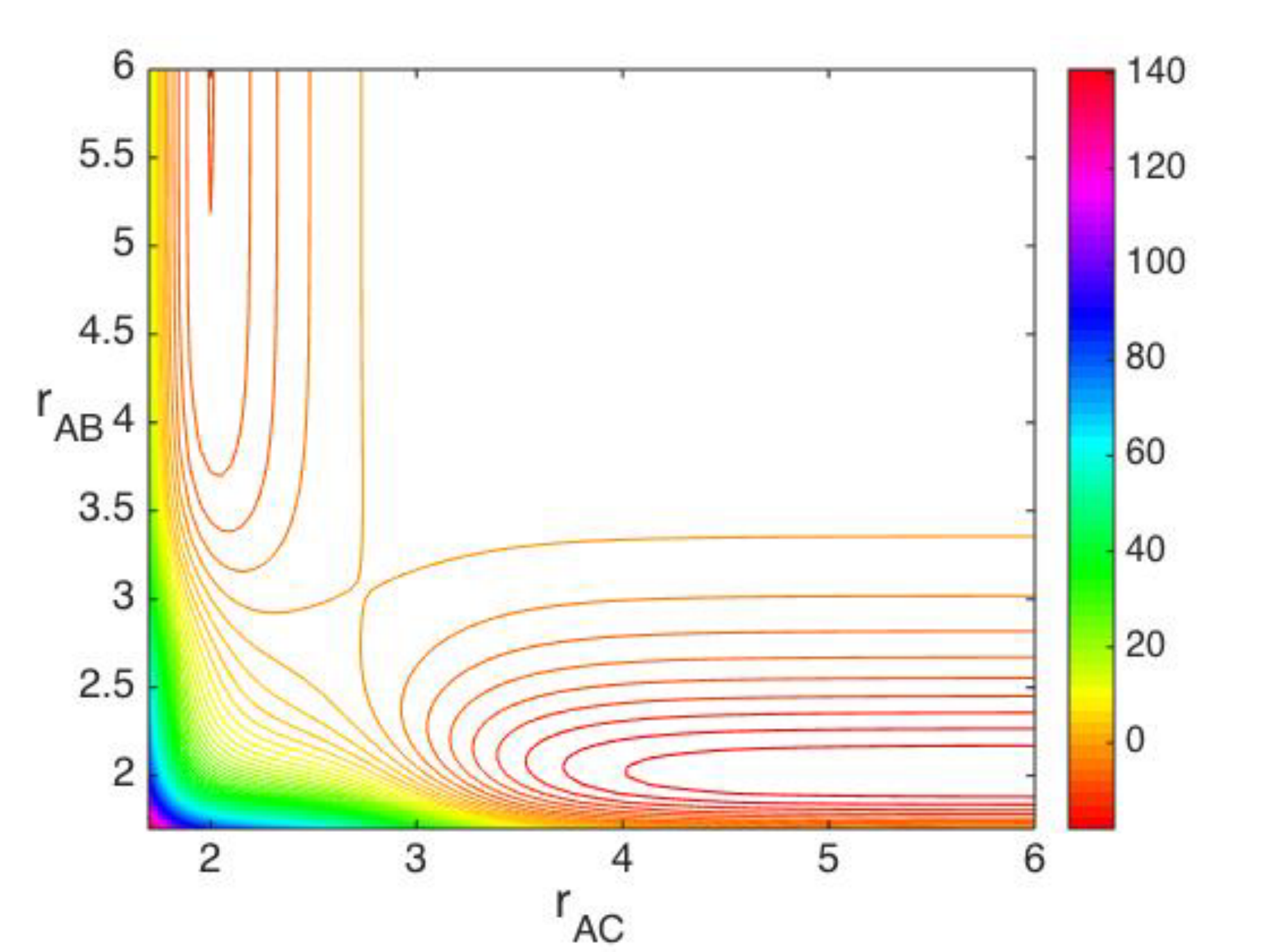}
	\caption{Potential energy level sets}
	\label{fig:biased-contour-plot}
\end{subfigure}
\begin{subfigure}[t]{0.3\textwidth}
	\centering
	\includegraphics[width=\textwidth]{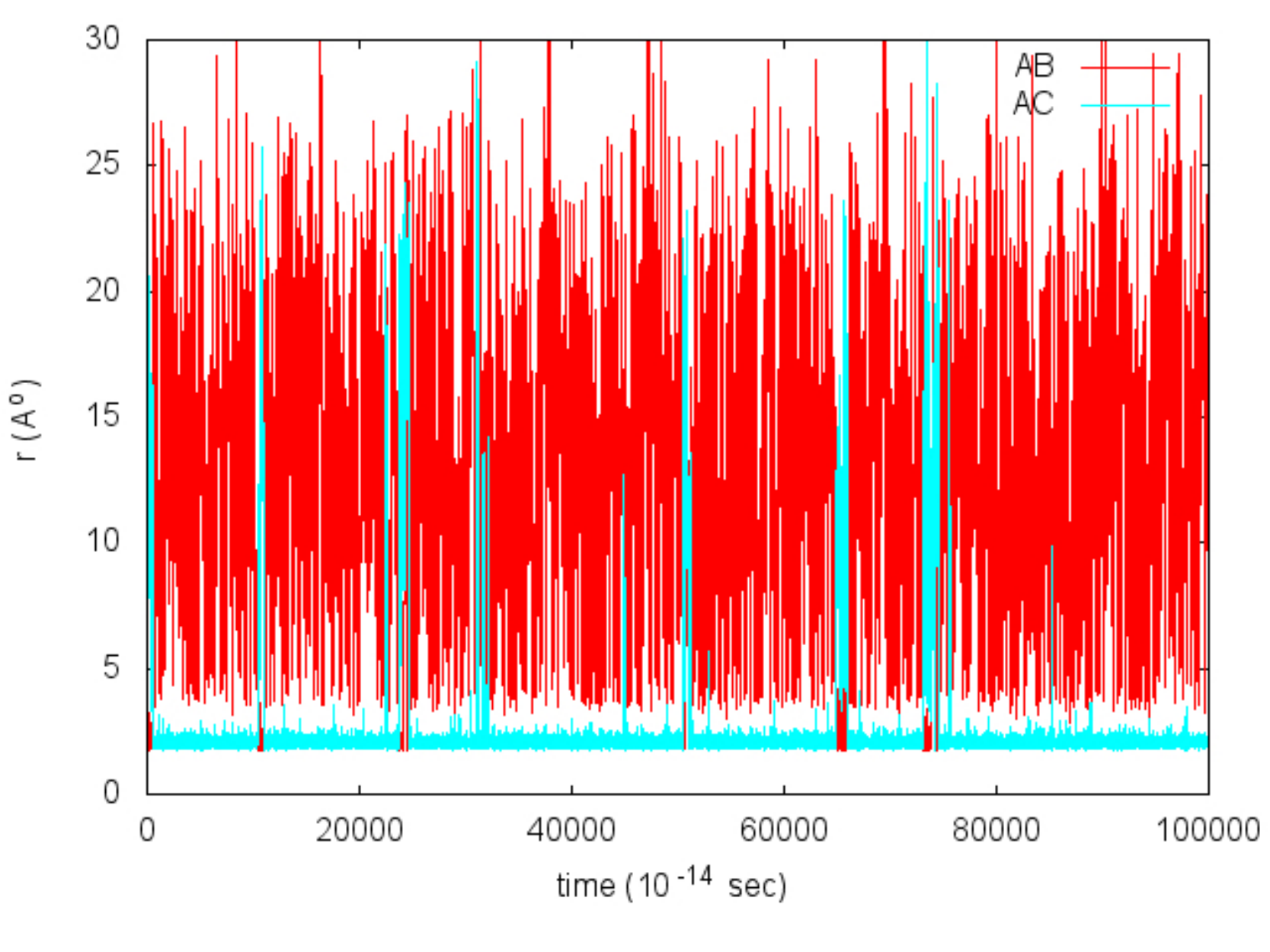}
	\caption{Relative distances between active sites on molecules \textbsf{A}, \textbsf{B} (red) and molecules \textbsf{A}, \textbsf{C} (cyan)}
	\label{fig:biased-relative-distances}
\end{subfigure}
\caption{Simulation of biased (2:1 well-depth), bond breaking chemical reaction, \eqref{eq:3-molecule-chemical-reaction}.
(a) The potential energy \eqref{eq:numerical-encoding-potential} for the system. The parameters are given under simulation 2 in Supplementary Table II.
(b) The level set plot of the potential energy. 
(c) A typical trajectory of the simulation. The \cyan{cyan} trace denotes the distance between molecules \textbsf{A} and \textbsf{C} ($r_{\textbsf{AC}} = \norm{\bvec{x}_{2} - \bvec{x}_{5}} $), whereas the \red{red} trace corresponds to the distance between molecules \textbsf{A} and \textbsf{B} ($r_{\textbsf{AB}} = \norm{\bvec{x}_{1} - \bvec{x}_{3}} $). Initially, \textbsf{A} and \textbsf{B} are near their equilibrium distance (2 \angstrom) and \textbsf{C} is far from \textbsf{A}. We see a successful $\textbsf{AB} + \textbsf{C} \rightarrow \textbsf{AC}  + \textbsf{B}$ event happening very soon (\red{red} trace is close to the equilibrium distance, then becomes large; \cyan{cyan} trace is large, then becomes small). The trace exhibits the bias towards a stable \textbsf{AC} bond, since the cyan trace is close to equilibrium longer than the red trace.}
\label{fig:chemical-reaction-simulation-2}
\end{figure}

The parameters of the second simulation are chosen so that the reaction is biased in favor of \textbsf{AC}. With the chosen parameters ($D_{\textbsf{AC}}/ D_{\textbsf{AB}} = 2$), the \textbsf{AC} is twice as stable as \textbsf{AB}. If the noise in the system is strong enough, it is expected that \textbsf{AC} is formed twice as often as \textbsf{AB} is formed. Figure \ref{fig:chemical-reaction-simulation-2} shows the potential energy for this simulation (Fig.\ \ref{fig:biased-potential-energy-surface}), its corresponding level sets (Fig.\ \ref{fig:biased-contour-plot}), and a realization of the simulation (Fig.\ \ref{fig:biased-relative-distances}). In the energy surface plot and the level set plot, the symmetry of the potential is evident. The realization shown in Fig.\ \ref{fig:biased-relative-distances} starts with \textbsf{AB} near its equilibrium length ($2\angstrom$) with \textbsf{C} far from \textbsf{A}. In this particular realization \textbsf{AC} forms very quickly. Figure \ref{fig:biased-relative-distances} shows the bias towards the more stable \textbsf{AC}. The system spends most of its time with a stable \textbsf{AC} molecule with a relatively small amount of time with a stable \textbsf{AB} molecule. Thus, biased reactions can be captured in the framework. A movie of a part of the unbiased reaction simulation can be found in Supplementary Movie 2.

%%%%%%%%%%%%%%%%%%%%%%%%%%%%%%%%
%%% CHEMICAL REACTION : SIMULATION 1 - UNBIASED REACTION
%%%%%%%%%%%%%%%%%%%%%%%%%%%%%%%%

%%%%%%%%%%%%%%%%%%%%%%%%%%%%%%%%
%%% CHEMICAL REACTION : SIMULATION 2 - BIASED REACTION
%%%%%%%%%%%%%%%%%%%%%%%%%%%%%%%%

%%%%%%%%%%%%%%%%%%%%%%%%%%%%%%%%%%%%%%%
%%% EXAMPLE 3: DNA TRANSCRIPTION 
%%%%%%%%%%%%%%%%%%%%%%%%%%%%%%%%%%%%%%%
\subsection*{DNA transcription model}\label{subsec:dna-transcription}

\begin{figure}[htbp]
\centering
\includegraphics[width=0.5\columnwidth]{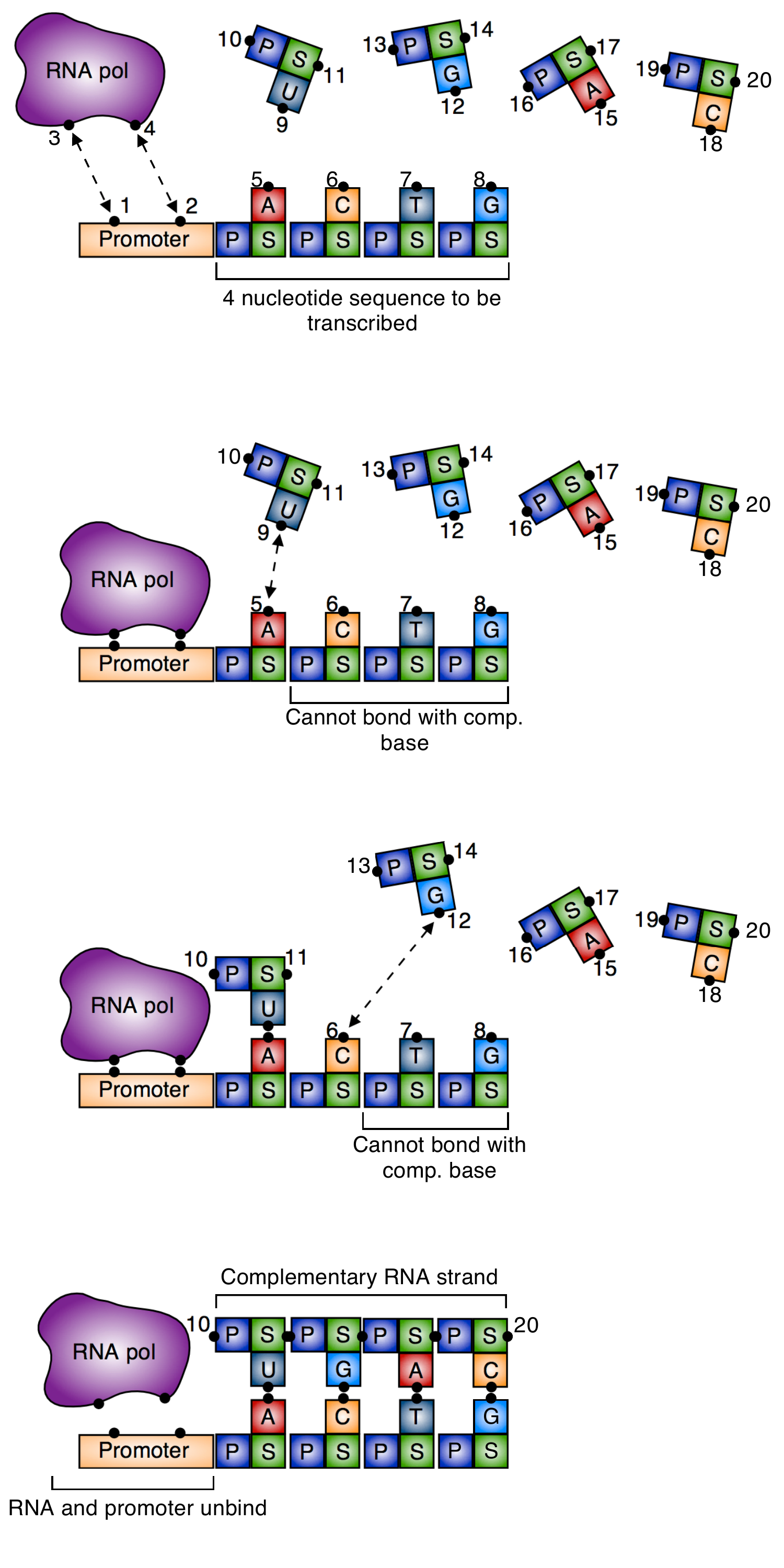}
\caption{Simple DNA transcription model. Free base nucleotides cannot bind with the DNA strand until RNA pol binds with the promoter. When RNA pol is bound to the promoter, the free nucleotides bind to the DNA strand \tsf{ACTG} sequentially from left to right. Once the complementary strand has formed, the RNA pol unbinds from the promoter and then the complementary strand can diffuse away. Once the RNA pol has diffused far enough away, the bonds between the complementary base pairs turn off and the complementary strand can diffuse away. Dashed arrows between sites denotes an active potential. The \tsf{P} blocks denote a phosphate group and the \tsf{S} blocks denote a sugar group.}
\label{fig:dna-trancription-model}
\end{figure}

The final example is inspired by DNA transcription \cite{Cramer:2000eb,Hahn:2004hw}. The model consists of a promoter region (sites 1 \& 2) to which RNA polymerase (RNA pol) binds (sites 3 \& 4), and a four nucleotide DNA strand, \tsf{ACTG}, to be transcribed (Fig.\ \ref{fig:dna-trancription-model}). As a first approximation of the transcription process, the movement of the RNA polymerase down the DNA chain and the unwinding/rewinding of the DNA have not been explicitly modeled. 

In the absence of the RNA pol, the free nucleotides cannot bind to their complementary nucleotides in the 4 nucleotide DNA strand (\tsf{ACTG} = (5,6,7,8)). Once RNA pol binds to the promoter, the first first nucleotide (\tsf{A}) in the DNA strand can bind to the free version of its complementary nucleotide (\tsf{U}). Before this binding happens, the remaining nucleotides in the strand (\tsf{CTG}) cannot bind with their (free) complementary nucleotides. Once \tsf{A} has bound to a free \tsf{U} nucleotide, the next nucleotide in the strand (\tsf{C}) can bind with a free \tsf{G} nucleotide, while the remaining two nucleotides (\tsf{TG}) still cannot bind with their complementary nucleotides. 
Once the free \tsf{G} has bound with \tsf{C}, the sugar and phosphate groups on \tsf{T} and \tsf{G} can bind to start forming the backbone of the complementary DNA strand.
This cascading process continues until each nucleotide in the original DNA strand \tsf{ACTG} has bound with its complementary nucleotide, resulting in the complementary RNA strand \tsf{UGAC}. At this point, the complementary strand and the RNA pol unbind from the original strand and promoter region, respectively. 

Supplementary Table III lists the reaction potentials for each of the interacting pairs. The nucleotide base pairs interact via a hydrogen bond $\phi_{\tsf{H}}$, whereas the sugar and phosphate groups covalently bond through $\phi_{\tsf{SP}}$. The interaction potentials for the system can be easily read from this table (see Supplementary Information Sec.\ IV.A).

Let us step through the rest of the logic in the order the reaction occurs:
\begin{compactenum}[(i)]
\item\label{item:rna-promoter-logic} The bonds between the RNA pol and the promoter region ($\Phi_{(1,3)}$ and $\Phi_{(2,4)}$) are ``on'' except when the complementary chain has formed and is still attached to the original base strand.
\item\label{item:a-t-logic} The \tsf{A-U} bond ($\Phi_{(5,9)}$) is ``on'' when the RNA pol has bonded with the promoter and the complementary chain has not formed. This second condition prevents the complementary strand from reattaching to the original DNA strand once it has been formed. It is ``off'' otherwise. 
\item\label{item:c-g-logic} The \tsf{C-G} bond ($\Phi_{(6,12)}$) is ``on'' when all of following conditions are true: (1) RNA pol has bonded with the promotor, (2) the \tsf{A-U} bond has formed. It is ``off'' otherwise.
\item\label{item:sugar-phosphate-t-g-backbone} The sugar-phosphate group bond $\Phi_{(11,13)}$ turns ``on'' after the \tsf{C-G} bond has formed. It is ``off'' otherwise.
\item\label{item:t-a-logic} The \tsf{T-A} bond ($\Phi_{(7,15)}$) is ``on'' when all of the following conditions are true: (1) RNA pol has bonded with the promotor, (2) the \tsf{A-U} bond has formed, (3) the \tsf{C-G} bond has formed, and (4) the $(11,13)$ sugar-phosphate bond is formed. It is ``off'' otherwise.
\item\label{item:sugar-phosphate-g-a-backbone} The sugar-phosphate group bond $\Phi_{(14,16)}$ turns ``on'' when \tsf{T-A} bond has formed. It is ``off'' otherwise.
\item\label{item:g-c-logic} The \tsf{G-C} bond ($\Phi_{(8,18)}$) is ``on'' when all of the following conditions are true: the conditions in \eqref{item:t-a-logic} are true, the \tsf{T-A} bond has formed, and the (14,16) sugar phosphate bond has formed. It is ``off'' otherwise.
\item\label{item:sugar-phosphate-a-c-backbone} The sugar-phosphate group bond $\Phi_{(17,19)}$ turns ``on'' when \tsf{G-C} bond has formed. It is ``off'' otherwise.
\end{compactenum}

A global potential derived from the above logic rules is given in Eq.\ \eqref{eq:dna-transcription-potential}. The exact form of the logic functions $L_{\bvec{p}}(\x)$ and the associated smooth encoding functions $S_{\bvec{p}}(\x)$ comprising the potential are given in the Supplementary Information. The derivation of the potential is not difficult, but lengthy. We refer the reader the Supplementary Information Sec.\ 4 for the details.
	%% ------------ EQUATION ------------ %%
	\begin{align}
	U(\x) 
	&= \underbrace{S_{(1,3)}(\x) \Phi_{(1,3)}(\x) + S_{(2,4)}(\x) \Phi_{(2,4)}(\x)}_{\textrm{RNA pol/promoter binding}} \nonumber \\
	&\quad+ \underbrace{S_{(5,9)}(\x) \Phi_{(5,9)}(\x)}_{\textrm{complementary \tsf{A-U} bond}} 
	+ \underbrace{S_{(6,12)}(\x) \Phi_{(6,12)}(\x)}_{\textrm{complementary \tsf{C-G} bond}} 
	+ \underbrace{S_{(7,15)}(\x) \Phi_{(7,15)}(\x)}_{\textrm{complementary \tsf{T-A} bond}} 
	+ \underbrace{S_{(8,18)}(\x) \Phi_{(8,18)}(\x)}_{\textrm{complementary \tsf{G-C} bond}} \nonumber \\
	&\quad+ \underbrace{ S_{(11,13)}(\x)\Phi_{(11,13)}(\x) + S_{(14,16)}(\x)\Phi_{(14,16)}(\x) + S_{(17,19)}(\x)\Phi_{(17,19)}(\x) .}_{\textrm{sugar-phosphate backbone for complementary RNA strand}} \label{eq:dna-transcription-potential}
	%&\quad + \underbrace{ S_{(10,20)}(\x) \Phi_{(10,20)}(\x).}_{\textrm{sugar-phosphate ends of complementary strand}} 
	\end{align}
	%% ---------------- END  ---------------- %%

\begin{figure}[htbp]
\begin{center}
\includegraphics[width=0.6\columnwidth]{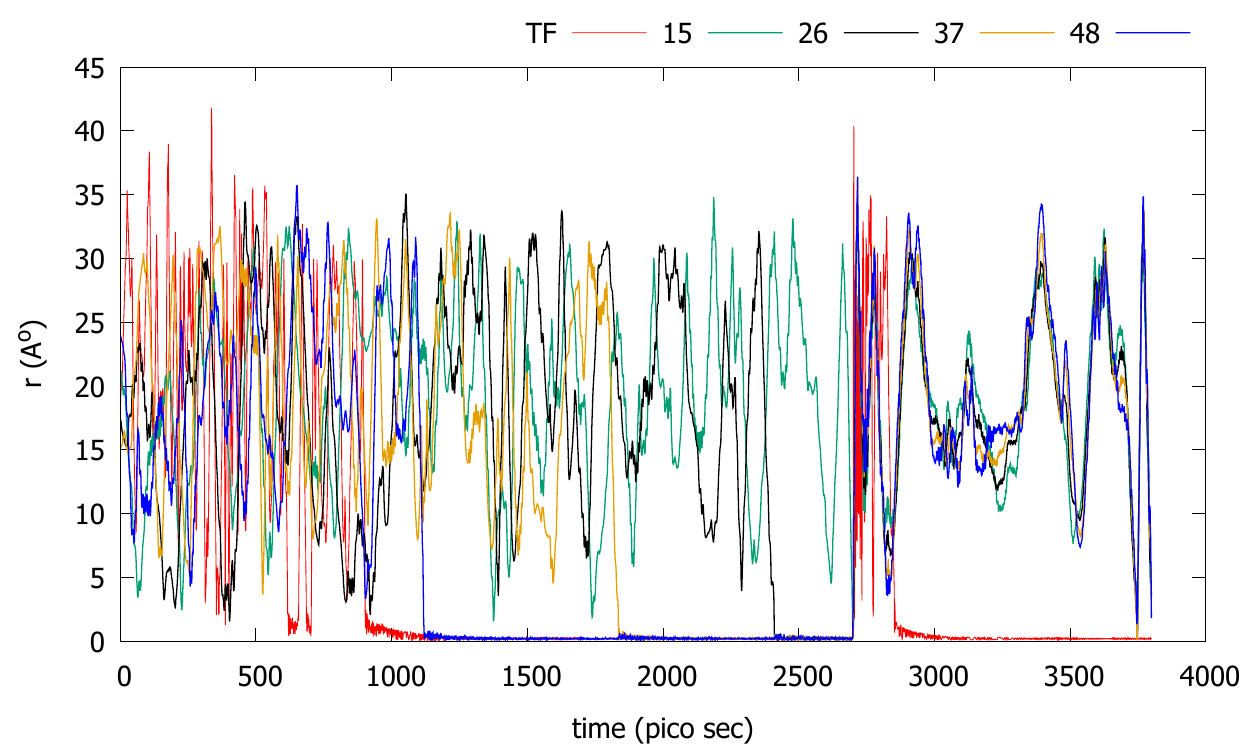}
\caption{DNA transcription. The red trace (TF) corresponds to the distance between the RNA pol and the promoter region. $r_{5,9}$ (teal), $r_{6,12}$ (black), $r_{7,15}$ (orange), and $r_{8,18}$ (blue) correspond to the sites on the complementary \tsf{A-U}, \tsf{C-G}, \tsf{T-A}, and \tsf{G-C} pairs from the base strand and the free nucleotides. At the start, the RNA pol and the free nucleotides diffuse around in space. Around 900 ps, the RNA pol binds to the promoter region (TF trace $\approx 0$). The free nucleotides then bind in the the order of the designed logic. \tsf{U} binds to \tsf{A} ($r_{5,9}\approx 0$) around 1100 ps; \tsf{G} binds with \tsf{C} ($r_{6,12} \approx 0$) between 1800 and 1900 ps; \tsf{A} binds to \tsf{T} ($r_{7,15} \approx 0$) around 2400 ps; and finally \tsf{C} binds to \tsf{G} ($r_{8,18} \approx 0$) around 2700 ps. Once this final free nucleotide has bounded with its complement, the complementary chain is finished formed and unbinds as does the RNA pol. The RNA pol can rebind to the promoter region, but the complementary strand cannot rebind to the original DNA strand.}
\label{fig:dna-transcription-trace}
\end{center}
\end{figure}

Figure \ref{fig:dna-transcription-trace} shows a trace of the pairwise distances between atoms for a typical simulation using this potential in LAMMPS (parameters in Supplementary Information Table IV). We use the same qualitative approximation of the force as was used in the inhibitor molecule example. For simplicity, all the potentials are taken to be Morse potentials. The red trace (TF) corresponds to the distance between the RNA pol and the promoter region. The variables $r_{5,9}$ (teal), $r_{6,12}$ (black), $r_{7,15}$ (orange), and $r_{8,18}$ (blue) correspond to the sites on the complementary \tsf{A-U}, \tsf{C-G}, \tsf{T-A}, and \tsf{G-C} pairs from the base strand and the free nucleotides. At the start, the RNA pol and the free nucleotides diffuse around in space. Around 900 ps, the RNA pol binds to the promoter region (TF trace $\approx 0$). The free nucleotides then bind in the the order of the designed logic. \tsf{U} binds to \tsf{A} ($r_{5,9}\approx 0$) around 1100 ps; \tsf{G} binds with \tsf{C} ($r_{6,12} \approx 0$) between 1800 and 1900 ps; \tsf{A} binds to \tsf{T} ($r_{7,15} \approx 0$) around 2400 ps; and finally \tsf{C} binds to \tsf{G} ($r_{8,18} \approx 0$) around 2700 ps. Once this final free nucleotide has bounded with its complement, the complementary chain has finished forming and unbinds as does the RNA pol. The RNA pol can rebind to the promoter region, but the complementary RNA strand cannot rebind to the original DNA strand. This is exactly the behavior designed into the potential. Supplementary Movie 3 in the Supplementary Information shows one simulation of the DNA transcription.

%%%%%%%%%%%%%%%%%%%%%%%%%%%%%%%%
%%% CONCLUSIONS
%%%%%%%%%%%%%%%%%%%%%%%%%%%%%%%%
\section*{Conclusions}

We have developed and demonstrated a methodology and mathematical framework for obtaining an approximate interaction potential for a system which respects known coarse-level behavior. This methodology develops a semi-empirical model for the system by encoding the known coarse-level physics into logic functions that then modify simple pairwise potentials. Each logic function's only role is to turn its associated pairwise potential on or off. A smooth multi-body interaction potential is obtained by replacing each logic function with a smoothed variant. The reader may wish to think of the resulting approximate potential as a linear combination of pairwise potentials where instead of the coefficients taking scalar values, they are encoding functions capturing the coarse-level logic.

Three relatively simple examples demonstrated our methodology: a simple inhibitor molecule mechanism, a chemical reaction with bond breaking, and a model inspired by DNA transcription. While these examples were simple and inspired by biophysical and chemistry applications, we stress that the methodology is quite general and not restricted to these application domains or only simple problems. Any system that is driven by a potential can utilize this methodology to its benefit.

The result of our procedure is the approximation of a complicated, high-dimensional potential with a lower-dimensional representation that still respects the relevant physics. A significant reduction in the dimensionality of the system is possible; instead of accounting for every interaction between a large number of components, we now only need as many variables as are needed to correctly model the coarse-level logic. In the bond breaking example, the potential capturing the logic was 8-dimensional, whereas the dimension of the configurations space was 12. The same system modeled at the quantum level is much more complicated. Since the bond breaking event is the relevant physics, the reduced order model is accurate enough for this purpose.

With this dimensional reduction, the ability to accurately simulate large, complicated systems within a computational design framework is feasible. The resultant models can be wrapped in an optimization loop as part of exploratory computational experiments, such as for the development of new drug therapies, or  as part of an engineering design loop. This in turn allows for the faster and cheaper development of new technologies and products.

We note that the developed framework can be potentially used in reverse: not for approximation to a given physical process with coarse-grained logic given, but  for design of molecular processes with logic prescribed by a designer. This is achieved by providing to the designer the specifications of molecules that can carry the logic out.

%\bibliography{references}

\begin{thebibliography}{10}
\expandafter\ifx\csname url\endcsname\relax
  \def\url#1{\texttt{#1}}\fi
\expandafter\ifx\csname urlprefix\endcsname\relax\def\urlprefix{URL }\fi
\providecommand{\bibinfo}[2]{#2}
\providecommand{\eprint}[2][]{\url{#2}}

\bibitem{Valastyan:2014gp}
\bibinfo{author}{Valastyan, J.~S.} \& \bibinfo{author}{Lindquist, S.}
\newblock \bibinfo{title}{{Mechanisms of protein-folding diseases at a
  glance}}.
\newblock \emph{\bibinfo{journal}{Disease Models {\&} Mechanisms}}
  \textbf{\bibinfo{volume}{7}}, \bibinfo{pages}{9--14} (\bibinfo{year}{2014}).
\newblock \urlprefix\url{http://dx.doi.org/10.1242/dmm.013474}.

\bibitem{VanDuin:2001ud}
\bibinfo{author}{van Duin, A. C.~T.}, \bibinfo{author}{Dasgupta, S.},
  \bibinfo{author}{Lorant, F.} \& \bibinfo{author}{Goddard~III, W.~A.}
\newblock \bibinfo{title}{{ReaxFF: a reactive force field for hydrocarbons}}.
\newblock \emph{\bibinfo{journal}{J. Phys. Chem. A}}
  \textbf{\bibinfo{volume}{105}}, \bibinfo{pages}{9396--9409}
  (\bibinfo{year}{2001}).
\newblock \urlprefix\url{http://dx.doi.org/10.1021/jp004368u}.

\bibitem{Friesner:2005cf}
\bibinfo{author}{Friesner, R.~A.}
\newblock \bibinfo{title}{{Ab initio quantum chemistry: Methodology and
  applications}}.
\newblock \emph{\bibinfo{journal}{PNAS}} \textbf{\bibinfo{volume}{102}},
  \bibinfo{pages}{6648--6653} (\bibinfo{year}{2005}).
\newblock \urlprefix\url{http://dx.doi.org/10.1073/pnas.0408036102}.

\bibitem{Aktulga:2012dy}
\bibinfo{author}{Aktulga, H.~M.}, \bibinfo{author}{Pandit, S.~A.},
  \bibinfo{author}{van Duin, A. C.~T.} \& \bibinfo{author}{Grama, A.~Y.}
\newblock \bibinfo{title}{{Reactive Molecular Dynamics: Numerical Methods and
  Algorithmic Techniques}}.
\newblock \emph{\bibinfo{journal}{SIAM J. Sci. Comput.}}
  \textbf{\bibinfo{volume}{34}}, \bibinfo{pages}{C1--C23}
  (\bibinfo{year}{2012}).
\newblock \urlprefix\url{http://dx.doi.org/10.1137/100808599}.

\bibitem{Gillespie:1992gb}
\bibinfo{author}{Gillespie, D.~T.}
\newblock \bibinfo{title}{{A rigorous derivation of the chemical master
  equation}}.
\newblock \emph{\bibinfo{journal}{Physica A: Statistical Mechanics and its
  Applications}} \textbf{\bibinfo{volume}{188}}, \bibinfo{pages}{404--425}
  (\bibinfo{year}{1992}).
\newblock \urlprefix\url{http://dx.doi.org/10.1016/0378-4371(92)90283-V}.

\bibitem{Kiel}
\bibinfo{author}{Kiel, C.}, \bibinfo{author}{Yus, E.} \&
  \bibinfo{author}{Serrano, L.}
\newblock \bibinfo{title}{Engineering signal transduction pathways}.
\newblock \emph{\bibinfo{journal}{Cell}} \textbf{\bibinfo{volume}{140}},
  \bibinfo{pages}{33--47} (\bibinfo{year}{2010}).
\newblock \urlprefix\url{http://dx.doi.org/10.1016/j.cell.2009.12.028}.

\bibitem{Laub}
\bibinfo{author}{Laub, M.} \& \bibinfo{author}{Goulian, M.}
\newblock \bibinfo{title}{Specificity in two-component signal transduction
  pathways.}
\newblock \emph{\bibinfo{journal}{Annual Review of Genetics}}
  \textbf{\bibinfo{volume}{41}}, \bibinfo{pages}{121--145}
  (\bibinfo{year}{2007}).
\newblock
  \urlprefix\url{http://dx.doi.org/10.1146/annurev.genet.41.042007.170548}.

\bibitem{lnui}
\bibinfo{author}{Inui, M.}, \bibinfo{author}{Martello, G.} \&
  \bibinfo{author}{Piccolo, S.}
\newblock \bibinfo{title}{Microrna control of signal transduction}.
\newblock \emph{\bibinfo{journal}{Nat Rev Mol Cell Biol}}
  \textbf{\bibinfo{volume}{11}}, \bibinfo{pages}{252--263}
  (\bibinfo{year}{2010}).
\newblock \urlprefix\url{http://dx.doi.org/10.1038/nrm2868}.

\bibitem{yarden}
\bibinfo{author}{Yarden, Y.} \& \bibinfo{author}{Sliwkowski, M.~X.}
\newblock \bibinfo{title}{Untangling the erbb signalling network}.
\newblock \emph{\bibinfo{journal}{Nature Reveiws Molecular Cell Biology}}
  \textbf{\bibinfo{volume}{2}}, \bibinfo{pages}{127--137}
  (\bibinfo{year}{2001}).
\newblock \urlprefix\url{http://dx.doi.org/10.1038/35052073}.

\bibitem{sako}
\bibinfo{author}{Sako, Y.}, \bibinfo{author}{Minoghchi, S.} \&
  \bibinfo{author}{Yanagida, T.}
\newblock \bibinfo{title}{Single-molecule imaging of egfr signalling on the
  surface of living cells}.
\newblock \emph{\bibinfo{journal}{Nat Cell Biol}} \textbf{\bibinfo{volume}{2}},
  \bibinfo{pages}{168--172} (\bibinfo{year}{2000}).
\newblock \urlprefix\url{http://dx.doi.org/10.1038/35004044}.

\bibitem{berger1}
\bibinfo{author}{Berger, B.}, \bibinfo{author}{Shor, P.~W.},
  \bibinfo{author}{Tucker-Kellogg, L.} \& \bibinfo{author}{King, J.}
\newblock \bibinfo{title}{Local rule-based theory of virus shell assembly}.
\newblock \emph{\bibinfo{journal}{Proceedings of the National Academy of
  Sciences}} \textbf{\bibinfo{volume}{91}}, \bibinfo{pages}{7732--7736}
  (\bibinfo{year}{1994}).
\newblock \urlprefix\url{http://www.pnas.org/content/91/16/7732.abstract}.

\bibitem{berger2}
\bibinfo{author}{Berger, B.}, \bibinfo{author}{King, J.},
  \bibinfo{author}{Schwartz, R.} \& \bibinfo{author}{Shor, P.}
\newblock \bibinfo{title}{Local rule mechanism for selecting icosahedral shell
  geometry}.
\newblock \emph{\bibinfo{journal}{Discrete Applied Mathematics}}
  \textbf{\bibinfo{volume}{104}}, \bibinfo{pages}{97 -- 111}
  (\bibinfo{year}{2000}).
\newblock \urlprefix\url{http://dx.doi.org/10.1016/S0166-218X(00)00187-6}.

\bibitem{schwartz}
\bibinfo{author}{Schwartz, R.}, \bibinfo{author}{Shor, P.~W.},
  \bibinfo{author}{Prevelige, P.~E.} \& \bibinfo{author}{Berger, B.}
\newblock \bibinfo{title}{Local rules simulation of the kinetics of virus
  capsid self-assembly}.
\newblock \emph{\bibinfo{journal}{Biophysical journal}}
  \textbf{\bibinfo{volume}{75}}, \bibinfo{pages}{2626--2636}
  (\bibinfo{year}{1998}).
\newblock \urlprefix\url{http://dx.doi.org/10.1016/S0006-3495(98)77708-2}.

\bibitem{klavins1}
\bibinfo{author}{Klavins, E.}, \bibinfo{author}{Ghrist, R.} \&
  \bibinfo{author}{Lipsky, D.}
\newblock \bibinfo{title}{A grammatical approach to self-organizing robotic
  systems}.
\newblock \emph{\bibinfo{journal}{Automatic Control, IEEE Transactions on}}
  \textbf{\bibinfo{volume}{51}}, \bibinfo{pages}{949--962}
  (\bibinfo{year}{2006}).
\newblock \urlprefix\url{http://dx.doi.org/10.1109/TAC.2006.876950}.

\bibitem{klavins2}
\bibinfo{author}{Klavins, E.}
\newblock \bibinfo{title}{Programmable self-assembly}.
\newblock \emph{\bibinfo{journal}{Control Systems, IEEE}}
  \textbf{\bibinfo{volume}{27}}, \bibinfo{pages}{43 --56}
  (\bibinfo{year}{2007}).
\newblock \urlprefix\url{http://dx.doi.org/10.1109/MCS.2007.384126}.

\bibitem{Whitesides}
\bibinfo{author}{Whitesides, G.~M.} \& \bibinfo{author}{Grzybowski, B.}
\newblock \bibinfo{title}{Self-assembly at all scales}.
\newblock \emph{\bibinfo{journal}{Science}} \textbf{\bibinfo{volume}{295}},
  \bibinfo{pages}{2418--2421} (\bibinfo{year}{2002}).
\newblock \urlprefix\url{http://dx.doi.org/10.1126/science.1070821}.

\bibitem{Whitesides2}
\bibinfo{author}{Whitesides, G.~M.} \& \bibinfo{author}{Boncheva, M.}
\newblock \bibinfo{title}{Beyond molecules: Self-assembly of mesoscopic and
  macroscopic components}.
\newblock \emph{\bibinfo{journal}{PNAS}} \textbf{\bibinfo{volume}{99}},
  \bibinfo{pages}{4769--4774} (\bibinfo{year}{2002}).
\newblock \urlprefix\url{http://dx.doi.org/10.1073/pnas.082065899}.

\bibitem{yurii}
\bibinfo{author}{Vlasov, Y.~A.}, \bibinfo{author}{Bo, X.-Z.},
  \bibinfo{author}{Sturm, J.~C.} \& \bibinfo{author}{Norris, D.~J.}
\newblock \bibinfo{title}{On-chip natural assembly of silicon photonic bandgap
  crystals}.
\newblock \emph{\bibinfo{journal}{Nature}} \textbf{\bibinfo{volume}{414}},
  \bibinfo{pages}{289--293} (\bibinfo{year}{2001}).
\newblock \urlprefix\url{http://dx.doi.org/10.1038/35104529}.

\bibitem{Whitesides3}
\bibinfo{author}{Gracias, D.~H.}, \bibinfo{author}{Tien, J.},
  \bibinfo{author}{Breen, T.~L.}, \bibinfo{author}{Hsu, C.} \&
  \bibinfo{author}{Whitesides, G.~M.}
\newblock \bibinfo{title}{Forming electrical networks in three dimensions by
  self-assembly}.
\newblock \emph{\bibinfo{journal}{Science}} \textbf{\bibinfo{volume}{289}},
  \bibinfo{pages}{1170--1172} (\bibinfo{year}{2000}).
\newblock \urlprefix\url{http://dx.doi.org/10.1126/science.289.5482.1170}.

\bibitem{licata}
\bibinfo{author}{Licata, N.~A.} \& \bibinfo{author}{Tkachenko, A.~V.}
\newblock \bibinfo{title}{Errorproof programmable self-assembly of
  dna-nanoparticle clusters}.
\newblock \emph{\bibinfo{journal}{Phys. Rev. E}} \textbf{\bibinfo{volume}{74}},
  \bibinfo{pages}{041406} (\bibinfo{year}{2006}).
\newblock \urlprefix\url{http://dx.doi.org/10.1103/PhysRevE.74.041406}.

\bibitem{vale}
\bibinfo{author}{Valentine, M.~T.} \& \bibinfo{author}{Gilbert, S.~P.}
\newblock \bibinfo{title}{To step or not to step? how biochemistry and
  mechanics influence processivity in kinesin and eg5}.
\newblock \emph{\bibinfo{journal}{Current Opinion in Cell Biology}}
  \textbf{\bibinfo{volume}{19}}, \bibinfo{pages}{75--81}
  (\bibinfo{year}{2007}).
\newblock
  \urlprefix\url{http://www.sciencedirect.com/science/article/pii/S095506740600192X}.

\bibitem{gunjan_prog}
\bibinfo{author}{Thakur, G.~S.}
\newblock \emph{\bibinfo{title}{Encoding Information in Coarse Grain Models for
  Self-Assembling Systems}}.
\newblock Ph.D. thesis, \bibinfo{school}{University of California, Santa
  Barbara} (\bibinfo{year}{2011}).

\bibitem{Konnov:2008dy}
\bibinfo{author}{Konnov, A.~A.}
\newblock \bibinfo{title}{{Remaining uncertainties in the kinetic mechanism of
  hydrogen combustion}}.
\newblock \emph{\bibinfo{journal}{Combustion and Flame}}
  \textbf{\bibinfo{volume}{152}}, \bibinfo{pages}{507--528}
  (\bibinfo{year}{2008}).
\newblock \urlprefix\url{http://dx.doi.org/10.1016/j.combustflame.2007.10.024}.

\bibitem{Hong:2011dj}
\bibinfo{author}{Hong, Z.}, \bibinfo{author}{Davidson, D.~F.} \&
  \bibinfo{author}{Hanson, R.~K.}
\newblock \bibinfo{title}{{An improved H2/O2 mechanism based on recent shock
  tube/laser absorption measurements}}.
\newblock \emph{\bibinfo{journal}{Combustion and Flame}}
  \textbf{\bibinfo{volume}{158}}, \bibinfo{pages}{633--644}
  (\bibinfo{year}{2011}).
\newblock \urlprefix\url{http://dx.doi.org/10.1016/j.combustflame.2010.10.002}.

\bibitem{lammps}
\bibinfo{author}{Plimpton, S.}
\newblock \bibinfo{title}{Fast parallel algorithms for short-range molecular
  dynamics}.
\newblock \emph{\bibinfo{journal}{J. Comput. Phys.}}
  \textbf{\bibinfo{volume}{117}}, \bibinfo{pages}{1--19}
  (\bibinfo{year}{1995}).
\newblock \urlprefix\url{http://dx.doi.org/10.1006/jcph.1995.1039}.

\bibitem{Jensen:2007wr}
\bibinfo{author}{Jensen, F.}
\newblock \emph{\bibinfo{title}{{Introduction to Computational Chemistry}}}
  (\bibinfo{publisher}{John Wiley {\&} Sons}, \bibinfo{year}{2007}),
  \bibinfo{edition}{second} edn.

\bibitem{Cramer:2000eb}
\bibinfo{author}{Cramer, P.} \emph{et~al.}
\newblock \bibinfo{title}{{Architecture of RNA Polymerase II and Implications
  for the Transcription Mechanism}}.
\newblock \emph{\bibinfo{journal}{Science}} \textbf{\bibinfo{volume}{288}},
  \bibinfo{pages}{640--649} (\bibinfo{year}{2000}).
\newblock \urlprefix\url{http://dx.doi.org/10.1126/science.288.5466.640}.

\bibitem{Hahn:2004hw}
\bibinfo{author}{Hahn, S.}
\newblock \bibinfo{title}{{Structure and mechanism of the RNA polymerase II
  transcription machinery}}.
\newblock \emph{\bibinfo{journal}{Nat Struct Mol Biol}}
  \textbf{\bibinfo{volume}{11}}, \bibinfo{pages}{394--403}
  (\bibinfo{year}{2004}).
\newblock \urlprefix\url{http://dx.doi.org/10.1038/nsmb763}.

\bibitem{arnold}
\bibinfo{author}{Arnold, V.}
\newblock \emph{\bibinfo{title}{Mathematical methods of classical Mechanics}},
  vol.~\bibinfo{volume}{60} of \emph{\bibinfo{series}{Graduate Text in
  Mathematics}} (\bibinfo{publisher}{Springer-Verlag}, \bibinfo{year}{1989}),
  \bibinfo{edition}{2} edn.

\bibitem{Katznelson:2002uy}
\bibinfo{author}{Katznelson, Y.}
\newblock \emph{\bibinfo{title}{{An Introduction To Harmonic Analysis}}}.
\newblock Cambrdige Mathematical Library (\bibinfo{publisher}{Cambridge
  University Press}, \bibinfo{year}{2002}), \bibinfo{edition}{3} edn.

\bibitem{Lin:2007tp}
\bibinfo{author}{Lin, H.} \& \bibinfo{author}{Truhlar, D.~G.}
\newblock \bibinfo{title}{{QM/MM: what have we learned, where are we, and where
  do we go from here?}}
\newblock \emph{\bibinfo{journal}{Theoretical Chemistry Accounts}}
  (\bibinfo{year}{2007}).
\newblock \urlprefix\url{http://dx.doi.org/10.1007/s00214-006-0143-z}.

\end{thebibliography}

%%%%%%%%% END BIBLIO %%%%%%%%% 

\section*{Acknowledgements}

Ryan Mohr and Igor Mezi\'{c} received funding from Army Research Office grant ARO-MURI W911NF-14-1-0359.

\section*{Author contributions statement}

The original conceptualization of the idea is due to G.S.T and I.M. The precise formulation of the concept is due to  G.S.T, I.M., and R.M. The details of mathematical formulation is due to R.M. and numerical simulations were done by G.S.T. The majority of the writing of the manuscript was done by R.M. with G.S.T contributing. All authors discussed the results and commented on the manuscript at all stages.

\section*{Additional information}

\textbf{Competing financial interests} The authors declare no competing financial interests.

\end{document}